\documentstyle[preprint,aps,tighten,epsfig]{revtex}
\begin{document}
\draft
\title{ A DIRAC-HARTREE-BOGOLIUBOV APPROXIMATION \\
FOR FINITE NUCLEI }
\author{B.V. Carlson}
\address{Departamento de F\'{\i}sica, \\
Instituto Tecnol\'ogico da Aeron\'autica, \\
Centro T\'ecnico Aeroespacial, 12228-900 S\~ao Jos\'e dos Campos,\\
S\~ao Paulo, Brazil}
\author{D. Hirata\footnote{present address: GANIL, BP5027, Bd 
Becquerel, 14076 Caen Cedex 5, France}}
\address{Japan Synchrotron Radiation Research Institute (JASRI) - SPring-8, \\
Kamigori, Hyogo, 678-12, Japan }
\date{\today }
\maketitle

\begin{abstract}
We develop a complete Dirac-Hartree-Fock-Bogoliubov approximation to the
ground state wave function and energy of finite nuclei. We apply it to
spin-zero proton-proton and neutron-neutron pairing within the
Dirac-Hartree-Bogoliubov approximation (we neglect the Fock term), using a
zero-range approximation to the relativistic pairing tensor. We study the
effects of the pairing on the properties of the even-even nuclei of the
isotopic chains of Ca, Ni and Sn (spherical) and Kr and Sr
(deformed), as well as the $N$=28 isotonic chain, and compare
our results with experimental data and with other recent calculations.
\end{abstract}


\narrowtext

\section{Introduction}

The advances made at radioactive nuclear beam facilities provide us with an
increasing amount of information on nuclei far away from the stability line.
We now have access to experimental measurements of the masses, radii and
deformations of unstable nuclei in a wider region of the nuclear chart than
ever before. Studies of exotic nuclei have revealed new features, such as
neutron halos\cite{Tan85} and neutron skins\cite{Tan92,Suz95} and bring new
perspectives to nuclear physics\cite{Ham89}. Planned facilities around the
world plan to study unstable nuclei up to the r-process region or even
beyond. This will enable us to investigate where and how exotic phenomena of
nuclear structure appear in the region far from the stability line\cite
{Tan95,RNB4}.

In recent years relativistic many-body theories have been applied to nuclei
and nuclear matter with remarkable success\cite{Ser86,Rei89,Ser92}. The
relativistic Brueckner-Hartree-Fock (RBHF) theory has been shown capable of
reproducing the saturation properties of nuclear matter using interaction
parameters that describe the two-nucleon bound state and 
scattering\cite{Bro90}. Its
phenomenological version, the relativistic mean field model (RMF), has been
successfully applied to the description of many of the ground-state
properties of stable and unstable nuclei\cite{Mil72,Bro78,Bro81,Hor81,Pri87,%
Bou87,Gam90,Hir91,Sha92,Hir93,She93,Lal95,Rin96,Pom97} and has been shown
capable of simultaneously reproducing the properties of stable and
unstable nuclei over a wide mass range of the periodic table\cite{Hir97}.

The relativistic mean field approach has also been applied to describe the
structure of very exotic nuclei. In these calculations, the pairing
interaction has been neglected or simply treated by a nonrelativistic BCS
type of approximation\cite{Hir97,Lal98}. For nuclei on or near the stability
line, the BCS approach provides a reasonably good description of pairing
properties. However, for drip-line nuclei, the Fermi level is close to the
continuum and the coupling between bound and continuum states should be taken
into account explicitly. The pairing correlations and the mean fields should
be calculated simultaneously in order to obtain an accurate description of
the ground state properties of drip-line nuclei.

A more precise relativistic description of pairing correlations, a
Dirac-Hartree-Fock-Bogoliubov (DHFB) approximation, was developed some
time ago\cite{Bai84} and, more recently, applied to nuclear matter
calculations\cite{Kuc91,Gui96,Mat97,Car97}. The first calculations in
nuclear matter furnished pairing gaps much larger than those obtained in
nonrelativistic calculations using realistic interactions.\cite{Kuc91}
The results suggested that the meson-exchange interactions adjusted
to describe nuclear matter saturation might not be adequate for the
particle-particle pairing channel. However, these calculations did
not perform a complete self-consistent DHFB calculation. Rather, they
combined a relativistic Hartree mean field calculation with a
nonrelativistic calculation of the pairing field using the nonrelativistic
reduction of the meson-exchange potential. Self-consistent calculations
including the full Dirac structure of the self-energy and pairing fields
resulted in more reasonable values for the pairing gap.\cite{Gui96,Mat97}
Yet these DHFB calculations, using various sets of meson-exchange
parameters that furnish similar good descriptions of nuclear saturation,
still resulted in values of the pairing gap that were not consistent with
nonrelativistic calculations nor consistent among themselves.
In Ref. \onlinecite{Car97}, these discrepancies were resolved by associating
the pairing gap of each set of meson-exchange parameters with the set's
description of
low-energy two-nucleon scattering in the $^1$S$_0$ channel. When each of the
parameter sets was supplemented with a momentum cutoff, so that they each
described the two-nucleon $^1$S$_0$ virtual state correctly, they all provided 
consistent values for the pairing gap that very close to the nonrelativistic
ones, at low densities. Thus, the conclusion of Ref.~\onlinecite{Kuc91} was 
found to be correct afterall. The meson-exchange interactions adjusted
to describe nuclear matter saturation are indeed inadequate for the pairing
channel, when they do not describe the two-nucleon $^1$S$_0$ channel well.
In the case of nuclear matter, however, we have found that this inadequacy
can be easily remedied.

Even after the different sets of meson-exchange parameters have been augmented
so as to describe the two-nucleon $^1$S$_0$ channel, they still differ
slightly among themselves at densities greater than about one-eight of the
saturation density, where mean field effects begin to play a role. Most
importantly, they all differ in a consistent manner from the gap functions
obtained using realistic nonrelativistic two-nucleon interactions. The
relativistic pairing gap consistently falls
to zero at lower density (by about one-half) of that for which it
disappears in nonrelativistic calculations. Much as the Dirac structure of the
nuclear mean field weakens the effective interaction as the density
increases, leading to saturation, so the Dirac structure of the pairing
field  weakens the effective pairing interaction with density, leading to
the suppression of the pairing field relative to the nonrelativistic one.
An estimate of the relativistic effects on quasi-deuteron pairing shows that
the same effects
also suppress quasi-deuteron pairing at the saturation density\cite{Elg98}.
We thus expect a density-dependent suppression of the pairing field to be a
general feature when we take the Dirac structure of the field into account.

The relativistic Hartree mean field + nonrelativistic pairing field 
approximation of Ref.~\onlinecite{Kuc91} can and has been extended to
finite nuclei. This hybrid relativistic Hartree-Bogoliubov (RHB)
approximation, using a non-relativistic finite-range Gogny interaction
in the pairing channel, has been applied extensively, and quite successfully,
to the calculation of the ground state properties of
spherical\cite{Men96,Gon96,Pos97,Lal98a,Lal98b,Men98,Vre98,Lal98c,Vre98b} and
deformed nuclei\cite{Lal98d}. We would expect reasonable results from such
a model as long as the mean field is well described by the meson exchange
interaction and the $^1$S$_0$ two-nucleon channel is well described by the
effective pairing interaction. This is indeed the case in the references
cited here. However, we might expect its pairing interaction to be 
less reliable than a meson-exchange one when extrapolated far beyond the
stability valley, the condition for which its parameters have been adjusted.

Aside from simply being more pleasing aesthetically, we are motivated to
retain the Dirac structure of the pairing interaction in our calculations
by the success of our studies of $^1$S$_0$ pairing in nuclear matter and
by their differences from the nonrelativistic results. The success of our
nuclear matter calculations lead us to believe that a good description of
nuclei can be obtained using a Dirac pairing interaction based on a
meson-exchange one. Such a description would be of general interest, since the
Dirac structure of a meson-exchange interaction could be extended
beyond the valley of stability with better reliability than could a
nonrelativistic interaction. A nonrelativistic effective interaction must
incorporate the density dependence furnished by the Dirac structure of a
meson-exchange interaction into its effective parameters. This density
dependence changes as we consider nuclei further and further from the
stability valley, thus limiting the range of application of a effective
interaction adjusted to describe the conditions in the valley.    

The differences between our results and nonrelativistic ones in nuclear
matter suggest that interesting effects of the Dirac structure of the pairing
interaction might appear even in nuclei within the stability valley. 
The suppression of pairing in the saturated interior region, in DHFB
calculations, similar to that observed in nuclear matter,
would result in greater localization of the pairing field on the nuclear
surface than is the case in nonrelativistic or RHB calculations. This, in
turn, would tend to make the nucleus more rigid and tend to diminish its
deformation.
We then would expect to observe a smaller pairing field in DHB calculations
that furnish the same deformation as nonrelativistic or RHB ones, or a smaller
deformation in DHB calculations that furnish the same value of the pairing
field as nonrelativistic or RHB ones. 

Here we present a fully relativistic Dirac-Hartree-Bogoliubov (DHB)
approximation for axially deformed nuclei. We use a direct extension of the
DHFB approximation developed in Ref.\cite{Gui96}. We neglect the exchange
term in the self-energy but retain the Dirac structure of the pairing
interaction and pairing field. The neglect of the exchange term seems
a reasonable approximation here, as we restrict our calculations to
the exchange of $\sigma$, $\rho$ and $\omega$ mesons, for which the
exchange effects can be fairly well accomodated by adjustments in the
coupling constants. The characteristics of the pairing field and of its Dirac
structure will form the center of our study. Our goal is to extend our
successful DHFB description of nuclear matter to a DHB description of 
finite nuclei.

We use a local (zero-range) approximation to the meson-exchange interaction
in the pairing channel. A zero-range approximation to the pairing
interaction can be justified by an analysis of the
length scales in a typical DHB calculation. Taking oscillator wave functions
as a guide, we can estimate the wavelength in one dimension of a wave
function of quantum number $n$ (in that dimension) as $\lambda_n\approx 4.4
b_0/\sqrt{n}$, where $b_0$ is the oscillator length, $b_0\approx A^{1/6}$
fm, and $A$ is the mass number of the nucleus. In the fairly extreme case of
the N=16 shell of $^{16}$O, we find a wavelength of $\lambda_{16}\approx 1.75$
fm. The wavelength is larger in heavier nuclei or in lower shells.
On the other hand, the range of a typical meson-exchange interaction
is $r_x=\hbar c/m$, where $m$ is the meson mass. For the lightest of the
mesons that we include, the $\sigma$ meson, we find a range of
$r_{\sigma}<0.5$ fm, substantially smaller than the wavelength $\lambda_{16}$.
Even the term of longest range of the Gogny interaction has
$r_2=1.2$ fm only. Thus we find that the interaction range is typically
smaller than the length scale of the most energetic level, making the
effects due to the finite range of the interaction small. This analysis has,
in fact, been confirmed numerically by Meng\cite{Men98}, who found
zero-range and finite-range Gogny RHB calculations to yield almost
identical results.

In Section I of the following, we derive the self-consistency equations for
the self-energy and pairing fields. In Section II, we discuss the properties
of static solutions to the equations and, in Section III, reduce these to
the form in which we use them. In Section IV, we briefly describe the numerical
method used in the calculations and, in Section V, we discuss the results of
the calculations. We summarize and conclude in Section VI.

\section{The Mean Field Equations for $\Sigma$ and $\Delta$}

\label{sec2}

In this section we present the Lagrangian of the model and the covariant
equations for the self-energy and pairing mean fields, $\Sigma$ and $\Delta$.

We designate by $\psi (x)$, $\sigma (x)$, $\omega^{\mu }(x)$,
$\vec{\rho}^{\mu }(x)$, and $A^{\mu }(x)$, the field operators at the point $x$
associated to the nucleons and mesons $\sigma $, $\omega $, $\rho $, and $%
\gamma $ respectively. The quantum numbers $(J^{\pi },T)$ for each meson
with spin $J$, intrinsic parity $\pi $, and isospin $T$ are
\[
\sigma (0^{+},0),\qquad \omega (1^{-},0),\qquad \rho (1^{-},1),\qquad \gamma
(1^{-},-).
\]
We designate the effective meson-nucleon coupling constants by $g_{s}$,$%
\,\,g_{v}\,$and $g_{\rho }$ and the respective bare masses by $m_{s}$,$%
\,\,m_{v}\,$and $m_{\rho }$. The interaction of the massless $\gamma $ does
not conserve isospin, coupling to the protons alone with coupling constant $%
e $. (We will ignore the interaction with the anomalous magnetic moments of
the nucleons). The nucleon bare mass is $M$ and we assume in the present
model that the nucleons and mesons are point-like. We will not take the $\pi$
meson into account, although its effects could be important in the
exchange terms we will consider. We will assume that these
effects can be described through the effective couplings of the mesons
we do include. These assumptions are typical of the simplest meson-exchange
models of nuclear structure.

The Lagrangian density is given by
\[
{\cal L}={\cal L}_{0}+{\cal L}_{int}\;,
\]
where ${\cal L}_{0}$ is the free Lagrangian density \vspace{0.2cm}
\begin{eqnarray}
{\cal L}_{0}(x)&=&\overline{\psi }(x)[i\rlap\slash\partial -M]\psi (x)
+\frac{1}{2}\partial _{\mu }\sigma (x)\partial ^{\mu }\sigma (x)
-U(\sigma(x))-\frac{1}{4}F_{\mu \nu }F^{\mu \nu }  \nonumber \\
& & \qquad -\frac{1}{4}\Omega _{\mu \nu }\Omega ^{\mu \nu }
+\frac{1}{2}m_{v}^{2}\omega_{\mu }(x)\omega^{\mu }(x)
-\frac{1}{4}\vec{G}_{\mu \nu }\cdot\vec{G}^{\mu \nu }
+\frac{1}{2} m_{\rho }^{2}\vec{\rho}_{\mu }(x)\cdot\vec{\rho}^{\mu }(x)\,,
\end{eqnarray}
with vector field tensors
\begin{eqnarray}
F_{\mu \nu }&=&\partial _{\mu }A_{\nu }-\partial _{\nu }A_{\mu }\,,  \nonumber
\\
\Omega _{\mu \nu }&=&\partial _{\mu }\omega_{\nu }
-\partial _{\nu }\omega_{\mu }\,,
\nonumber \\
\vec{G}_{\mu \nu }&=&\partial _{\mu }\vec{\rho}_{\nu }-
\partial _{\nu }\vec{\rho}_{\mu }\,,  \nonumber
\end{eqnarray}
and a nonlinear $\sigma$ potential,
\begin{equation}
U(\sigma(x))=\frac{1}{2}m_s^2 \sigma(x)^2+\frac{1}{3}g_3\sigma(x)^3
+\frac{1}{4}g_4\sigma(x)^4\,.
\end{equation}
The baryon spinor $\psi(x)$ has four Dirac components for each of two
isospin projections -- $m_t=1/2$ for protons and $m_t=-1/2$ for
neutrons -- for a total of eight components. We have included the cubic
and quartic terms of the scalar field $\sigma(x)$
in the free Lagrangian density as we will only consider their contributions
to the scalar mean field. For this purpose, we may formally include
them in the `free' scalar meson propagator, when it is convenient to
do so.

We take the interaction terms in the Lagrangian density to have the
simplest possible form consistent with their Lorentz and isospin structure,
\begin{eqnarray}
{\cal L}_{int}(x)=g_{s}\overline{\psi }(x)\sigma (x)\psi (x) &-&
g_{v}\overline{\psi }(x)\gamma _{\mu }\omega^{\mu }(x)\psi (x)
-\frac{1}{2}g_{\rho }\overline{\psi }(x)\gamma _{\mu }
\vec{\tau}\cdot \vec{\rho}^{\mu }(x)\psi (x)
\nonumber \\
&&-e\overline{\psi }(x)\frac{\left( 1+\tau _{3}\right) }{2}
\gamma _{\mu}A^{\mu }(x)\psi (x).
\end{eqnarray}
In particular, we will not consider tensor couplings of the vector mesons.

We wish to characterize the average effect of the interactions of a nucleon
with the other nucleons through an effective single-particle Lagrangian, $%
L_{eff}$ , given in terms of the two fields, $\Sigma $ and $\Delta $. The
self-energy $\Sigma $ describes the average interaction of a nucleon with
the surrounding matter. The pairing field $\Delta $ and its conjugate $\bar{%
\Delta}$ describe, respectively, the formation and destruction of pairs
during the propagation. In particular, the definition of $\Delta $ makes use
of correlated pairs of time-reversed single-particle states, in agreement
with the original idea of Cooper \cite{Coo56}. Generalizing slightly the
development due to Gorkov\cite{Gor58}, we
introduce such pairs by using an extended form of the time-reversed states,
which we designate by $\psi _{T}$. Designating the time reversal operator by
${\cal T}$, the usual time-reversed conjugate $\psi ^{({\cal T})}$ of the
Dirac field operator $\psi $ is given by \cite{Itz80}
\[
\psi ^{({\cal T})}(x)={\cal T}\psi (x){\cal T}^{-1}=B\overline{\psi }^{T}(%
\tilde{x})=\gamma _{0}B\psi ^{*}(\tilde{x})\qquad
\]
where
\[
\tilde{x}=(-t,\vec{x})\,,\qquad \qquad \qquad B=\gamma _{5}C\,,
\]
and $C$ is the charge conjugation operator. We define $\psi _{T}$ as
\[
\psi _{T}(x)=\tau _{2}\otimes \,\psi ^{({\cal T})}(\tilde{x})\equiv A\,%
\overline{\psi }^{T}(x)\;,
\]
where $A=\tau _{2}\otimes B$ and $\tau _{2}$ is the antisymmetric Pauli
matrix, which operates here in the isospin space. Note that $A=A^{T}$ and $%
A^{*}=A^{\dagger }$. We then use the following ansatz for the effective
single-particle Lagrangian
\begin{eqnarray}
\int \!dt\,L_{eff} &=&\int \!d^{4}x\,d^{4}y\left\{ \overline{\psi }(x)\left[
i\rlap\slash\partial -M+\gamma _{0}\mu \right] \delta (x-y)\psi (y)-%
\overline{\psi }(x)\Sigma (x,y)\psi (y)\right.  \label{langeff} \\
&+&\left. \frac{1}{2}\overline{\psi }(x)\Delta (x,y)\psi _{T}(y)+\frac{1}{2}%
\overline{\psi _{T}}(x)\bar{\Delta}(x,y)\psi (y)\right\} ,  \nonumber
\end{eqnarray}
where $\delta (x-y)$ is a four-dimensional Dirac delta function and
\begin{equation}
\mu = \left(
\begin{array}{ll}
\mu _{p} & 0 \\
0 & \mu _{n}
\end{array}
\right)  \label{lagmult}
\end{equation}
is the isospin matrix of chemical potentials, which will be used as Lagrange
multipliers to fix the average number of protons and neutrons.

The symmetries of the effective mean-field Lagrangian under transposition and
Hermitian conjugation yield the following properties of the mean fields,
\begin{equation}
\Delta (x,y)=-A\;\Delta ^{T}\!(x,y)\;A^{\dagger }\;=-A\;(\Delta
\!(y,x))^{T}\;A^{\dagger }\qquad \mbox{and}\qquad \bar{\Delta}(x,y)=-A\;\bar{%
\Delta}^{T}\!(x,y)\;A^{\dagger },  \label{symm}
\end{equation}
and
\begin{equation}
\Sigma (x,y)=\gamma _{0}\Sigma ^{\dagger }(x,y)\gamma _{0}\qquad \qquad %
\mbox{and}\qquad \qquad \Delta (x,y)=\gamma _{0}\bar{\Delta}^{\dagger
}(x,y)\gamma _{0}\;.  \label{herm}
\end{equation}
The first of these symmetry conditions requires that
the pair wave function be antisymmetric under exchange while the
second guarantees real energy eigenvalues and probability conservation.

We can put the effective Lagrangian $L_{eff}$ in a more symmetrical form by
noting that
\begin{eqnarray}
\int \!d^{4}xd^{4}y&&\overline{\psi }(x)\left[ {}\right. (i\rlap\slash%
\partial -M+\gamma _{0}\mu )\delta (x-y)-\Sigma (x,y)\left. {}\right] \psi
(y)  \label{langeffsym} \\
&=&\int \!d^{4}xd^{4}y\;\;\overline{\psi _{T}}(x)\left[ (i\rlap\slash\partial
+M-\gamma _{0}\mu )\delta (x-y)+\Sigma _{T}(x,y)\right] \psi _{T}(y)\,,
\nonumber
\end{eqnarray}
where
\begin{equation}
\Sigma _{T}(x,y)=A\Sigma ^{T}(x,y)A^{\dagger }\;,
\end{equation}

The effective Lagrangian can then be rewritten in matrix form as
{\footnotesize
\begin{eqnarray}
\int \!dtL_{eff} &=& \frac{1}{2}\int d^{4}xd^{4}y\;\;(\overline{\psi }(x),%
\overline{\psi }_{T}(x))  \label{hfblang} \\
&&\times\left(
\begin{array}{cc}
(i\rlap\slash\partial -M+\gamma _{0}\mu )\delta (x-y)-\Sigma (x,y) & \Delta
(x,y) \\
\bar{\Delta}(x,y) & (i\rlap\slash\partial +M-\gamma _{0}\mu )\delta (x-y)
+\Sigma _{T}(x,y)
\end{array}
\right) \left(
\begin{array}{c}
\psi (y) \\
\psi _{T}(y)
\end{array}
\right) \,,  \nonumber
\end{eqnarray}
} which immediately yields the following coupled equations of motion for the
fields $\psi $ and $\psi _{T}$, which we will call the Dirac-Gorkov
equation, {\footnotesize
\begin{eqnarray}
\int d^{4}y\left(
\begin{array}{cc}
(i\rlap\slash\partial -M+\gamma _{0}\mu )\delta (x-y)-\Sigma (x,y) & \Delta
(x,y) \\
\bar{\Delta}(x,y) & (i\rlap\slash\partial +M-\gamma _{0}\mu )\delta (x-y)
+\Sigma _{T}(x,y)
\end{array}
\right) \left(
\begin{array}{c}
\psi (y) \\
\psi _{T}(y)
\end{array}
\right) = 0 \,.  \label{dirgor1}
\end{eqnarray}
}

Defining the generalized baryon field operator as
\[
\Psi (x)\,=\left(
\begin{array}{c}
\psi (x) \\
\psi _{T}(x)
\end{array}
\right) \;,
\]
we obtain a generalized baryon (quasi-particle) propagator
\begin{equation}
S(x,y)=\left(
\begin{array}{cc}
G(x,y) & F(x,y) \\
\tilde{F}(x,y) & \tilde{G}(x,y)
\end{array}
\right) =-i\left\langle \left(
\begin{array}{c}
\psi (x) \\
\psi _{T}(x)
\end{array}
\right) (\overline{\psi }(y)\ ,\ \overline{\psi }_{T}(y))\right\rangle \,,
\label{prop}
\end{equation}
where, by $\left\langle ...\right\rangle $, we mean the time-ordered
expectation value in the interacting nuclear ground state,
$\left\langle \widetilde{0}\right| T(...)\left| \widetilde{0}\right\rangle $.
We assume that the state $\left| \widetilde{0}\right\rangle $ contains only
nucleons interacting through the exchange of virtual mesons and contains no
real mesons.

We observe that $G(x,y)$ is the usual baryon propagator while $\tilde{G}%
(x,y) $ describes the propagation of baryons in time-reversed states. The
off-diagonal terms of $S(x,y)$ describe the propagation of correlated
baryons and are the relativistic generalizations of the anomalous
propagators defined by Gorkov \cite{Gor58}.

To derive the mean field equations, we first rewrite the interaction terms
of the Lagrangian density, ${\cal L}_{int}$, as
\begin{eqnarray}
{\cal L}_{int}(x)=-\sum_j\overline{\psi }(x)\Gamma _{j\alpha }(x)\phi
_j^\alpha (x)\psi (x)\;,  \label{lint}
\end{eqnarray}
where the Greek letters $\alpha ,\beta ,\cdots $ represent any indices
necessary for the correct description of the meson propagation and coupling
(Lorentz indices, isospin, etc.). The index $j$ indicates the mesons of the
model: $\sigma $, $\omega $, $\pi $, and $\rho $, while their respective
fields and meson-nucleon couplings are designated by the $\phi _j^\alpha (x)
$ and the $\Gamma _{j\alpha }(x)$.

We then rewrite the meson fields $\phi_j$ in terms of their sources as
\begin{eqnarray}
\phi_j^\alpha (x)=\int \!d^4yD_j^{\alpha \beta }(x-y)\overline{\psi }%
(y)\Gamma_{j\beta }(y)\psi (y)\;,  \label{meson}
\end{eqnarray}
where $D_j^{\alpha \beta }(x-y)$ is the Feynman propagator of meson $j$.
Here, we have included the nonlinear $\sigma$-meson terms in the
$\sigma$-meson propagator, so that they do not appear as corrections to
the source term.  Substituting in Eq.~(\ref{lint}) and
inserting a factor of 1/2 for reasons of symmetry, we have
\begin{eqnarray}
\int \!dt\,L_{int}=-\frac{1}{2}\sum_j\int d^4xd^4y\,\overline{\psi }%
(x)\Gamma _{j\alpha }(x)\psi (x)D_j^{\alpha \beta }(x-y)\overline{\psi }%
(y)\Gamma _{j\beta }(y)\psi (y)\,.  \label{eq:b11a}
\end{eqnarray}
Following Gorkov \cite{Gor58}, we then obtain the mean field contribution of
this interaction term by replacing each of the possible pairs of fermion
fields by its vacuum expectation value,
\begin{eqnarray}
\int dt\,(L_{int})_{eff}=-\frac{1}{2}\sum_j\int d^4xd^4yD_j^{\alpha\beta
}(x-y) &&\left\{ 2\overline{\psi }(x)\Gamma _{j\alpha }(x)\psi
(x)\left\langle \overline{\psi }(y)\Gamma _{j\beta }(y)\psi (y)\right\rangle
\right.  \label{eq:ii4} \\
&+&2\overline{\psi }(x)\Gamma _{j\alpha }(x)\left\langle \psi (x)\overline{%
\psi }(y)\right\rangle \Gamma _{j\beta }(y)\psi (y)  \nonumber \\
&-&\overline{\psi }(x)\Gamma _{j\alpha }(x)\left\langle \psi (x)\psi
^T(y)\right\rangle \Gamma _{j\beta }^T(y)\overline{\psi }^T(y)  \nonumber \\
&-&\left. \psi ^T(x)\Gamma _{j\alpha }^T(x)\left\langle \overline{\psi }^T(x)%
\overline{\psi }(y)\right\rangle \Gamma _{j\beta }(y)\psi (y)\,\right\} .
\nonumber
\end{eqnarray}
where $\left\langle ...\right\rangle $ is again the time-ordered expectation
value in the interacting nuclear-matter ground state.

The first term in this expression is a Hartree one, the second
a Fock exchange term while the last two, after using the definition of $\psi
_{T}$ to replace the transposed $\psi $'s, can be recognized as pairing
terms. Comparing the mean field contributions to those of the effective
quasi-particle Lagrangian, we can express the self-energy and pairing fields
in terms of the two-fermion vacuum expectation values as
\begin{eqnarray}
\Sigma (x,y) &=&\delta (x-y)\sum_{j}\Gamma _{j\alpha }(x)\int
d^{4}z\,D_{j}^{\alpha \beta }(x-z)\left\langle \overline{\psi }(z)\Gamma
_{j\beta }(z)\psi (z)\right\rangle \\
&&\qquad \qquad +\sum_{j}\Gamma _{j\alpha }(x)D^{\alpha \beta
}(x-y)\left\langle \psi (x)\overline{\psi }(y)\right\rangle \Gamma _{j\beta
}(y)\,,  \nonumber
\end{eqnarray}
and
\begin{eqnarray}
\Delta (x,y)=\sum_{j}\Gamma _{j\alpha }(x)D_{j}^{\alpha \beta
}(x-y)\left\langle \psi (x)\overline{\psi }_{T}(y)\right\rangle A\Gamma
_{\beta }^{T}(y)A^{\dagger }\,,
\end{eqnarray}
while the equation for $\bar{\Delta}(x-y)$ can be obtained from the 
equation for $\Delta (x-y)$ using the
Hermiticity condition of Eq.~(\ref{herm}). These expressions become
self-consistency equations when we evaluate the expectation values by using
their relationship to the generalized baryon propagator, Eq.~(\ref{prop}),
which is itself a function of the mean fields. We find
\begin{eqnarray}
\Sigma (x,y) &=&-i\delta (x-y)\sum_{j}\Gamma _{j\alpha }(x)\int
d^{4}z\,D_{j}^{\alpha \beta }(x-z)\mbox{Tr}\left[ \Gamma _{j\beta
}(z)G(z,z^{+})\right]  \label{sigeq} \\
&&\qquad \qquad +i\sum_{j}\Gamma _{j\alpha }(x)D^{\alpha \beta
}(x-y)G(x,y)\Gamma _{j\beta }(y)\,,  \nonumber
\end{eqnarray}
and
\begin{eqnarray}
\Delta (x,y)\;=\;i\sum_{j}\;\Gamma _{j\alpha }(x)D_{j}^{\alpha \beta
}(x-y)F(x,y)A\Gamma _{j\beta }^{T}(y)A^{\dagger }\;.  \label{delteq}
\end{eqnarray}
The number of protons $Z$ and neutrons $N$ are the expectation values of the
baryon number operators, $\hat{N}=\overline{\psi }(x)\gamma _{0}(1\pm \tau
_{3})\psi (x)/2$, which we rewrite in terms of the generalized baryon
propagator \cite{Ser86}, as
\begin{eqnarray}
\left.
\begin{array}{l}
Z \\
N
\end{array}
\right\} =\int d^3 x \left\langle \overline{\psi }(x)\gamma _{0} \frac{(1\pm
\tau _{3})}{2}\psi (x)\right\rangle =-i\int d^{3}x\mbox{Tr}\left[ \gamma _{0}%
\frac{(1\pm \tau _{3})}{2}G(x,x^{+})\right].  \label{nop}
\end{eqnarray}
The Lagrange multipliers $\mu _{p}$ and $\mu _{n}$, given in Eq.~(\ref
{lagmult}), are determined by requiring that these equations yield the
desired values of $Z$ and $N$.

The Hamiltonian density operator is given by the $\hat{T}^{00}$ component
of the energy-momentum tensor,
\begin{eqnarray}
\hat{H}=\hat{T}^{00}= -{\cal L}+\frac{\partial {\cal L}}
{\partial (\partial_t\psi )} \partial_t\psi
+\sum_j\frac{\partial {\cal L}} {\partial (\partial_t\phi_j^\alpha )}
\partial_t\phi_j^\alpha \;.  \label{hamop}
\end{eqnarray}
Neglecting the retardation terms associated with the time derivatives of the
meson fields in the ground state expectation value of Eq.~(\ref{hamop}), the
energy density that results can be written as
\begin{eqnarray}
{\cal E}(x) = <\hat{H}> &=& i\mbox{Tr}\left[ (i\vec{\gamma}\cdot
\vec{\partial}-M)G(x,x^+) \right] - \frac{1}{6}\,g_3\,\sigma(x)^3
-\frac{1}{4}\,g_4\,\sigma(x)^4  \label{energy} \\
& &-\frac{i}{2}\int d^4y\mbox{Tr}
\left[ \Sigma(x,y)G(y,x^+) - \Delta (x,y)\tilde{F}(y,x) \right]\,.  \nonumber
\end{eqnarray}
The total energy is obtained by integrating this density over space.

\medskip

\section{Properties of static solutions}

\smallskip

We will develop a static, ground-state solution to the self-consistency
equations. We write the temporal Fourier transform of the full HFB
propagator as
\begin{eqnarray}
S(\vec{x},\vec{y};\omega )=\left(
\begin{array}{cc}
G(\vec{x},\vec{y};\omega ) & F(\vec{x},\vec{y};\omega ) \\
\tilde{F}(\vec{x},\vec{y};\omega ) & \tilde{G}(\vec{x},\vec{y};\omega )
\end{array}
\right) &=&\sum_{\alpha }\left(
\begin{array}{c}
U_{\alpha }(\vec{x}) \\
V_{\alpha }(\vec{x})
\end{array}
\right) \frac{1}{\omega -\varepsilon _{\alpha }+i\eta }\left(
\begin{array}{cc}
\overline{U}_{\alpha }(\vec{y}), & \overline{V}_{\alpha }(\vec{y})
\end{array}
\right)  \nonumber \\
&+&\sum_{\beta }\left(
\begin{array}{c}
U_{\beta }(\vec{x}) \\
V_{\beta }(\vec{x})
\end{array}
\right) \frac{1}{\omega +\varepsilon _{\beta }-i\eta }\left(
\begin{array}{cc}
\overline{U}_{\beta }(\vec{y}), & \overline{V}_{\beta }(\vec{y})
\end{array}
\right) .  \label{statg}
\end{eqnarray}
The components $U_{\alpha ,\beta }$ and $V_{\alpha ,\beta }$ are Dirac
spinors corresponding to the normal and time-reversed components,
respectively, of the positive-frequenc, $\varepsilon _{\alpha }$, and
negative-frequency, $\varepsilon _{\beta }$, solutions to the Dirac-Gorkov
equation {\footnotesize
\begin{eqnarray}
\int d^{3}y\left(
\begin{array}{cc}
\gamma _{0}((\varepsilon +\mu )\delta (\vec{x}-\vec{y})-h(\vec{x},\vec{y}))
& \Delta (\vec{x},\vec{y}) \\
\bar{\Delta}(\vec{x},\vec{y}) & ((\varepsilon -\mu )\delta (\vec{x}-\vec{y}%
)+h_{T}(\vec{x},\vec{y}))\gamma _{0}
\end{array}
\right) \left(
\begin{array}{c}
U(\vec{y}) \\
V(\vec{y})
\end{array}
\right) = 0\,,  \label{autoeq1}
\end{eqnarray}
} where we have introduced the single-particle hamiltonian, $h(\vec{x},\vec{y%
})$, given by
\begin{equation}
h(\vec{x},\vec{y})=(-i\vec{\alpha}\cdot \vec{\nabla}+\beta M)\delta (\vec{x}-%
\vec{y})+\beta \Sigma (\vec{x},\vec{y}),
\end{equation}
with
\begin{equation}
h_{T}(\vec{x},\vec{y})=Ah^{T}(\vec{x},\vec{y})A^{\dagger }\qquad \qquad %
\mbox{and}\qquad \qquad \ h(\vec{x},\vec{y})=h^{\dagger }(\vec{x},\vec{y}).
\end{equation}
After multiplying on the left by the matrix {\footnotesize
$\left(
\begin{array}{cc}
\gamma _{0} & 0 \\
0 & 1
\end{array}
\right) $}, the Dirac-Gorkov equation, Eq.~(\ref{autoeq1}), can be written in
Hamiltonian form as a Hermitian eigenequation, {\footnotesize
\begin{eqnarray}
\int d^{3}y\left(
\begin{array}{cc}
(\varepsilon +\mu )\delta (\vec{x}-\vec{y})-h(\vec{x},\vec{y}) & \bar{\Delta}%
^{\dagger }(\vec{x},\vec{y}) \\
\bar{\Delta}(\vec{x},\vec{y}) & (\varepsilon -\mu )\delta (\vec{x}-\vec{y}%
)+h_{T}(\vec{x},\vec{y})
\end{array}
\right) \left(
\begin{array}{c}
U(\vec{y}) \\
\gamma _{0}V(\vec{y})
\end{array}
\right) = 0\,.  \label{dgconj}
\end{eqnarray}
} We thus conclude that the eigenvalues $\varepsilon $ are real.

After multiplying the complex conjugate of Eq.~(\ref{autoeq1}) on the left by
the matrix $\gamma _{0} A \otimes \tau_1$, we can manipulate it into a form 
which is identical to the original equation, except for the sign of
$\varepsilon $. We may thus conclude that the solutions to the Dirac-Gorkov
equation occur in pairs with real eigenvalues of opposite sign and
eigenvectors of the form,
\begin{equation}
\varepsilon =\varepsilon _{\alpha }:\left(
\begin{array}{c}
U(\vec{y}) \\
V(\vec{y})
\end{array}
\right) ,\qquad \qquad \varepsilon =-\varepsilon _{\alpha }:\left(
\begin{array}{c}
\gamma _{0}AV^{*}(\vec{y}) \\
\gamma _{0}AU^{*}(\vec{y})
\end{array}
\right) .  \label{eigenvec}
\end{equation}
Using the above properties of the eigensolutions, we could rewrite the full
HFB propagator of Eq.~(\ref{statg}) as a sum over the positive eigenvalues 
$\varepsilon _{\alpha }$ alone. However, we will
continue to use the form given in Eq.~(\ref{statg}), distinguishing the
positive and negative frequency solutions through explicit reference to one
or the other.

In the frequency representation, the self-consistency equations take the
form
\begin{eqnarray}
\Sigma (\vec{x},\vec{y};\omega ) &=&-i\delta (\vec{x}-\vec{y})\sum_{j}\Gamma
_{j\alpha }(\vec{x})\int d^{3}z\,D_{j}^{\alpha \beta }(\vec{x}-\vec{z}%
;0)\int \frac{d\omega }{2\pi }\mbox{Tr}\left[ \Gamma _{j\beta }(\vec{z})G(%
\vec{z},\vec{z};\omega ^{+})\right]  \label{wsigeq} \\
&&\qquad \qquad +i\sum_{j}\int \frac{d\omega^{\prime} }{2\pi } \Gamma
_{j\alpha }(\vec{x})D^{\alpha \beta }(\vec{x}-\vec{y}; \omega
-\omega^{\prime} )G(\vec{x},\vec{y};\omega^{\prime})\Gamma _{j\beta } (\vec{y%
})\,,  \nonumber
\end{eqnarray}
and
\begin{equation}
\Delta (\vec{x},\vec{y};\omega )\;=\;i\sum_{j}\;\int \frac{d\omega^{\prime}
}{2\pi } \Gamma _{j\alpha }(\vec{x})D_{j}^{\alpha \beta }(\vec{x}-\vec{y};
\omega -\omega^{\prime})F(\vec{x},\vec{y};\omega^{\prime} ) A\Gamma _{j\beta
}^{T}(\vec{y})A^{\dagger }\;,
\end{equation}
where we have taken the vertices to be time independent and have assumed a
time dependence of the form $t-t^{\prime }$ in all other quantities. We
evaluate the equations in the static limit of the meson propagators,

\begin{equation}
D^{\alpha \beta }(\vec{x}-\vec{y};\omega )\longrightarrow \ D^{\alpha \beta
}(\vec{x}-\vec{y};0)\equiv \ D^{\alpha \beta }(\vec{x}-\vec{y}).
\end{equation}
We then have for the self-consistency equations
\begin{eqnarray}
\Sigma (\vec{x},\vec{y}) &=&-i\delta (\vec{x}-\vec{y})\sum_{j}\Gamma
_{j\alpha }(\vec{x})\int d^{3}z\,D_{j}^{\alpha \beta }(\vec{x}-\vec{z})\int
\frac{d\omega }{2\pi }\mbox{Tr}\left[ \Gamma _{j\beta }(\vec{z})G(\vec{z},%
\vec{z};\omega ^{+})\right] \\
&&\qquad \qquad +i\sum_{j}\Gamma _{j\alpha }(\vec{x})D^{\alpha \beta }(\vec{x%
}-\vec{y})\int \frac{d\omega }{2\pi }G(\vec{x},\vec{y};\omega )\ \Gamma
_{j\beta }(\vec{y})\,,  \nonumber
\end{eqnarray}
and
\[
\Delta (\vec{x},\vec{y})\;=\;i\sum_{j}\Gamma _{j\alpha }(\vec{x}%
)D_{j}^{\alpha \beta }(\vec{x}-\vec{y})\int \frac{d\omega }{2\pi }F(\vec{x},%
\vec{y};\omega )\ A\Gamma _{j\beta }^{T}(\vec{y})A^{\dagger }\;.
\]
Finally, we evaluate the frequency integrals by closing the contour in the
upper half-plane, yielding
\begin{eqnarray}
\Sigma (\vec{x},\vec{y}) &=&\delta (\vec{x}-\vec{y})\sum_{j}\Gamma _{j\alpha
}(\vec{x})\int d^{3}z\,D_{j}^{\alpha \beta }(\vec{x}-\vec{z}%
)\sum_{\varepsilon _{\gamma }<0}\bar{U}_{\gamma }(\vec{z})\Gamma _{j\beta }(%
\vec{z})U_{\gamma }(\vec{z})  \label{sigeqf} \\
&&\qquad \qquad -\sum_{j}\Gamma _{j\alpha }(\vec{x})D^{\alpha \beta }(\vec{x}%
-\vec{y})\sum_{\varepsilon _{\gamma }<0}U_{\gamma }(\vec{x})\bar{U}_{\gamma
}(\vec{y})\ \Gamma _{j\beta }(\vec{y})\,,  \nonumber
\end{eqnarray}
and
\begin{equation}
\Delta (\vec{x},\vec{y})\;=-\sum_{j}\Gamma _{j\alpha }(\vec{x})D_{j}^{\alpha
\beta }(\vec{x}-\vec{y})\sum_{\varepsilon _{\gamma }<0}U_{\gamma }(\vec{x})%
\bar{V}_{\gamma }(\vec{y})\ A\Gamma _{j\beta }^{T}(\vec{y})A^{\dagger }\;,
\label{paireqf}
\end{equation}
where the sum runs over the negative frequency solutions,
$\varepsilon _{\gamma }<0$, of Eq.~(\ref{autoeq1}).

The expressions for the number of protons and neutrons can be written
similarly as
\begin{eqnarray}
\left.
\begin{array}{l}
Z \\
N
\end{array}
\right\} =\int d^{3}x\sum_{\varepsilon _{\gamma }<0}U_{\gamma }^{\dagger }(%
\vec{x})\frac{(1\pm \tau _{3})}{2}U_{\gamma }(\vec{x}).  \label{nopf}
\end{eqnarray}
The expression for the energy can be reduced to
\begin{eqnarray}
E&=&\int d^{3}x\,\left(\sum_{\varepsilon _{\gamma }<0}
U_{\gamma }^{\dagger }(\vec{x}) (\varepsilon _{\gamma }+\mu )
U_{\gamma }(\vec{x}) -\frac{1}{6}\,g_3\,\sigma(\vec{x})^3
-\frac{1}{4}\,g_4\,\sigma(\vec{x})^4 \right)  \label{eneqf}\\
& &-\frac{1}{2} \int d^{3}x d^3 y\,\sum_{\varepsilon _{\gamma }<0}\left(
\overline{U}_{\gamma }^{\dagger}(\vec{x})\Sigma(\vec{x},\vec{y})
U_{\gamma }(\vec{y}) -\overline{U}_{\gamma }^{\dagger }(\vec{x})
\Delta(\vec{x},\vec{y}) V_{\gamma }(\vec{y})\right) \,,  \nonumber
\end{eqnarray}
Note that the last term is real due to the hermiticity of the
Dirac-Gorkov equation.

The summation over negative frequency solutions in Eqs.~(\ref{sigeqf})
through (\ref{eneqf}) takes into account the occupation of all states, both
those in the Fermi sea ( the surface of which is now diffuse, due to the
pairing) and those in the Dirac sea. These expressions would thus require
renormalization to yield finite results. Instead, we simply truncate the
sums by excluding the solutions corresponding to quasi-particle states in
the Dirac sea. This approximation has been found to give reasonable
results in nuclear matter\cite{Gui96}.

Before reducing the self-consistency equations, together with the
Dirac-Gorkov equation, Eq.~(\ref{autoeq1}), to a form in which they can be
solved, we observe that, with $B=\gamma _{5}C$, each of the pair of wave
vectors, {\footnotesize $%
\left(
\begin{array}{l}
U(\vec{x}) \\
V(\vec{x})
\end{array}
\right) $ }and {\footnotesize $\left(
\begin{array}{l}
BU^{*}(\vec{x}) \\
BV^{*}(\vec{x})
\end{array}
\right) $ }, possesses the time-reversed Dirac
structure of the other. In Appendix A, we show that when these two states
are equally occupied, then they satisfy
the same Dirac-Gorkov equation with the same eigenvalue. We will assume this
to be the case in the following.

\smallskip

\section{Reducing the self-consistency equations}

\smallskip

We begin by analyzing the isospin structure of the self-energy and
pairing fields under the assumption of pure proton-proton and
neutron-neutron pairing. In this case, the solutions to the Dirac-Gorkov
equation will be either purely proton particle-hole ones or neutron
particle-hole ones of the form

\[
\Psi _{p}=\left(
\begin{array}{l}
U_{p} \\
0 \\
0 \\
V_{p}
\end{array}
\right) \qquad \qquad \mbox{and}\qquad \qquad \Psi _{n}=\left(
\begin{array}{l}
0 \\
U_{n} \\
V_{n} \\
0
\end{array}
\right) ,
\]
where each of the elements in the column vectors are themselves 4-component
Dirac spinors. Substitution of these into the self-consistency
equations yields isospin structures of the mean fields of the form
\begin{equation}
\Sigma (\vec{x},\vec{y})=\left(
\begin{array}{ll}
\Sigma _{p}(\vec{x},\vec{y}) & 0 \\
0 & \Sigma _{n}(\vec{x},\vec{y})
\end{array}
\right) \qquad \mbox{and}\qquad \Delta (\vec{x},\vec{y})=\left(
\begin{array}{ll}
0 &\Delta _{p}(\vec{x},\vec{y}) \\
\Delta _{n}(\vec{x},\vec{y}) & 0
\end{array}
\right) ,
\end{equation}
where each of the elements of the matrices are themselves 4x4 matrices. The
isospin dependent Dirac-Gorkov equation thus decouples into independent
equations for neutrons and protons. In their Hamiltonian form, these are
\begin{equation}
\int d^{3}y\left(
\begin{array}{cc}
(\varepsilon +\mu _{t})\delta (\vec{x}-\vec{y})-h_{t}(\vec{x},\vec{y}) &
\bar{\Delta}_{t}^{\dagger }(\vec{x},\vec{y}) \\
\bar{\Delta}_{t}(\vec{x},\vec{y}) & (\varepsilon -\mu _{t})\delta (\vec{x}-%
\vec{y})+h_{t}(\vec{x},\vec{y})
\end{array}
\right) \left(
\begin{array}{c}
U_{t}(\vec{y}) \\
\gamma _{0}V_{t}(\vec{y})
\end{array}
\right) =0\,,  \label{ham}
\end{equation}
where we have written the respective Dirac hamiltonian operators as
\begin{equation}
h_{t}(\vec{x},\vec{y})=(-i\vec{\alpha}\cdot \vec{\nabla}+\beta M)\delta
(\vec{x}-\vec{y})+\beta \Sigma _{t}(\vec{x},\vec{y})\qquad \qquad t=p,n
\end{equation}
and the Lagrange multipliers as $\mu _{p}=\mu +\delta \mu $ and $\mu
_{n}=\mu -\delta \mu $. We have also made use of the fact that $\Sigma _{tT}(%
\vec{x},\vec{y})=\Sigma _{t}(\vec{x},\vec{y})$, so that $h_{tT}(\vec{x},\vec{%
y})=h_{t}(\vec{x},\vec{y}),$ since we have assumed invariance under time
inversion of the Dirac structure.

We want to obtain the coupled equations for the case of an axially-deformed
nucleus. We take the z-axis to be the symmetry axis and use cylindrical
coordinates ($x=r_{\bot}\cos \varphi$, $y=r_{\bot}\sin \varphi$ , and $z$).
Although the total angular momentum $j$ is no longer a good quantum number,
its projection along the symmetry axis $\Omega $, as well as the parity $\pi
$, (and, of course, the isospin projection, which we also denote by $t$) are
still good quantum numbers. We write each of the 4-component spinors of the
wave-function as
\begin{equation}
U_{t\gamma }(\vec{x})=\left(
\begin{array}{l}
uf_{t\gamma }(\vec{x}) \\
i\,ug_{t\gamma }(\vec{x})
\end{array}
\right) =\frac{1}{\sqrt{2\pi }}\left(
\begin{array}{l}
uf_{t\gamma }^{+}(r_{\bot},z)\,e^{i(\Omega _{\gamma }-1/2)\varphi } \\
uf_{t\gamma }^{-}(r_{\bot},z)\,e^{i(\Omega _{\gamma }+1/2)\varphi } \\
i\,ug_{t\gamma }^{+}(r_{\bot},z)\,e^{i(\Omega _{\gamma }-1/2)\varphi } \\
i\,ug_{t\gamma }^{-}(r_{\bot},z)\,e^{i(\Omega _{\gamma }+1/2)\varphi }
\end{array}
\right) ,
\end{equation}
and
\begin{equation}
\gamma _{0}V_{t\gamma }(\vec{x})=\left(
\begin{array}{l}
vf_{t\gamma }(\vec{x}) \\
i\,vg_{t\gamma }(\vec{x})
\end{array}
\right) =\frac{1}{\sqrt{2\pi }}\left(
\begin{array}{l}
vf_{t\gamma }^{+}(r_{\bot},z)\,e^{i(\Omega _{\gamma }-1/2)\varphi } \\
vf_{t\gamma }^{-}(r_{\bot},z)\,e^{i(\Omega _{\gamma }+1/2)\varphi } \\
i\,vg_{t\gamma }^{+}(r_{\bot},z)\,e^{i(\Omega _{\gamma }-1/2)\varphi } \\
i\,vg_{t\gamma }^{-}(r_{\bot},z)\,e^{i(\Omega _{\gamma }+1/2)\varphi }
\end{array}
\right) ,
\end{equation}
in which the functions $uf_{t\gamma }^{\pm }$, $ug_{t\gamma }^{\pm }$, $%
vf_{t\gamma }^{\pm }$, and $vg_{t\gamma }^{\pm }$ are real. For each
solution with positive value of the angular momentum projection, $\Omega
_{\gamma }$, we have a time-reversed solution with the same energy but a
negative value of the angular momentum projection, -$\Omega _{\gamma }$,
given by $BU_{t\gamma }^{*}(\vec{x})$ and
$\gamma _{0}BV_{t\gamma }^{*}(\vec{x})$.
The densities that enter the Hartree terms of the self-energy can then
be written as
\begin{eqnarray}
\rho _{s}(r_{\bot},z)&=&2\sum_{\omega _{t\gamma }<0,t,\Omega _{\gamma
}>0} U_{t\gamma}^{\dagger}\gamma_0 U_{t\gamma}, \nonumber  \\
\rho _{B}(r_{\bot},z)&=&2\sum_{\omega _{t\gamma }<0,t,\Omega _{\gamma
}>0} U_{t\gamma}^{\dagger} U_{t\gamma}, \label{rhoss} \\
\rho _{3}(r_{\bot},z)&=&2\sum_{\omega _{t\gamma }<0,t,\Omega _{\gamma
}>0}2m_t\, U_{t\gamma}^{\dagger} U_{t\gamma}, \nonumber \\
\rho _{c}(r_{\bot},z)&=&2\sum_{\omega _{t\gamma }<0,t,\Omega _{\gamma
}>0}(m_t+1/2)\, U_{t\gamma}^{\dagger} U_{t\gamma},  \nonumber
\end{eqnarray}
in which the sum over states of different parity is implicit in the sum over
states of different energy. The local Hartree contribution to the
self-energy may be written in terms of these densities as
\begin{eqnarray}
\beta \Sigma _{H}(\vec{x},\vec{y}) &=&\delta (\vec{x}-\vec{y})\int
d^{3}z\,\left( -\,\beta \,g_{\sigma }^{2}\,d_{\sigma }(\vec{x}-\vec{z}%
)\,\rho _{s}(\vec{z})+g_{\omega }^{2}\,d^0_{\omega }(\vec{x}-\vec{z})\,\rho
_{B}(\vec{z})\right. \\
&&\left. \qquad\qquad
+\left( \frac{g_{\rho }}{2}\right) ^{2}\,\tau _{3\,}d^0_{\rho }(\vec{x%
}-\vec{z})\,\rho _{3}(\vec{z})+e^{2}\frac{(1+\tau _{3})}{2}\,d^0_{\gamma }(%
\vec{x}-\vec{z})\,\rho _{c}(\vec{z})\right),  \nonumber
\end{eqnarray}
where the meson propagators have been reduced to
\begin{equation}
d^0_{j}(\vec{x}-\vec{z})=\frac{1}{4\pi }\frac{\exp \left( -m_{j}\left|
\vec{x}-\vec{y}\right| \right) }{\left| \vec{x}-\vec{y}\right| },
\label{prop0}
\end{equation}
with the exception of the $\sigma$-meson propagator, which is still
assumed to contain the nonlinear terms. It is customary to write the
Hartree contribution to the self-energy
in terms of the mean fields associated with each of the mesons, as
\begin{equation}
\beta \Sigma _{H}(\vec{x})=-\beta g_{\sigma}\,\sigma(\vec{x})
+g_{\omega}\,\omega^0(\vec{x}) + \frac{g_{\rho}}{2}\,\tau_3\,\rho^{00}(\vec{x})
+e\,\frac{(1+\tau_3)}{2}\,A^0(\vec{x})\,,
\end{equation}
with
\begin{eqnarray}
\omega^0(\vec{x})&=&g_{\omega}\int d^3 z\,d^0_{\omega}(\vec{x}-\vec{z})
\rho_B(\vec{z})\,, \nonumber \\
\rho^{00}(\vec{x})&=&\frac{g_{\rho}}{2}\int d^3 z\,d^0_{\rho}(\vec{x}-\vec{z})
\rho_3(\vec{z})\,, \nonumber \\
A^0(\vec{x})&=&e\int d^3 z\,d^0_{\gamma}(\vec{x}-\vec{z})
\rho_c(\vec{z})\,,  \\
\sigma(\vec{x})&=&g_{\sigma}\int d^3 z\,d_{\sigma}(\vec{x}-\vec{z})
\rho_s(\vec{z}) \nonumber \\
&=&\int d^3 z\,d^0_{\sigma}(\vec{x}-\vec{z}) \left( g_{\sigma}\rho_s(\vec{z})
-g_3\,\sigma(\vec{x})^2-g_4\,\sigma(\vec{x})^3\right)\,.\nonumber
\end{eqnarray}
In the last expression, we have written the $\sigma$ mean field in
terms of its free propagator, as given in Eq.~(\ref{prop0}), and
included the nonlinear terms explicitly. The mean meson fields
possess the same axial symmetry as the densities.

The Fock exchange term of the self-energy has the form
\begin{eqnarray}
\Sigma _{F}(\vec{x},\vec{y})=&&-\sum_{j}\Gamma _{j\alpha }(\vec{x})D^{\alpha
\beta }(\vec{x}-\vec{y}) \\
&&\times\sum_{\varepsilon _{t\gamma }<0,t,\Omega _{\gamma }>0}
\chi_t\chi_t^{\dagger}\left( U_{t\gamma }(\vec{x})U_{t\gamma }^{\dagger }(%
\vec{y})+BU_{t\gamma }^{*}(\vec{x})U_{t\gamma }^{T}(\vec{y})B^{\dagger
}\right) \gamma _{0}\Gamma _{j\beta }(\vec{y})\,,  \nonumber
\end{eqnarray}
where we have written the isospin dependence of the intermediate states
in terms of the isospinor $\chi_t$. As has been discussed in the 
literature (cf. Ref. \onlinecite{Ser86} or Ref.\onlinecite{Lal98d}), 
the most important effects of the Fock terms due to the exchange of the 
short-range $\sigma$, $\omega$, and $\rho$ mesons can be taken into account
by using adjusted Hartree terms. This can be better understood by looking
at the zero-range limit of the meson propagators, for which the Fock terms
reduce to linear combinations of the Hartree ones. Here, we retain the small
but finite range of the meson propagators, but approximate the self-energy as
the Hartree contribution alone and use the parameters appropriate for
this purpose.

The pairing field arises from an exchange term, similar in form to the Fock
term of the self-energy. We can write it as
\begin{eqnarray}
\Delta (\vec{x},\vec{y})=&&-\sum_j\Gamma _{j\alpha }(\vec{x})D_j^{\alpha
\beta }(\vec{x}-\vec{y}) \\
&&\times\sum_{\varepsilon _{t\gamma} <0,t,\Omega _{\gamma} >0} \chi_t\chi_{%
\bar t}\left( U_{t\gamma }(\vec{x})V_{t\gamma }^{\dagger }(\vec{y}%
)+BU_{t\gamma }^{*}(\vec{x})V_{t\gamma }^T(\vec{y})B^{\dagger }\right) \
\gamma _0A\Gamma _{j\beta }^T(\vec{y})A^{\dagger },  \nonumber
\end{eqnarray}
where $\chi_{{\bar p},{\bar n}}=\chi_{n,p}$ are the time-reversed isospinors.

We neglect the Coulomb contribution to the pairing field here and
approximate the contributions of the other mesons using the
zero-range (infinite-mass) limit of the meson propagators. We also neglect the
contribution of the non-linear $\sigma$-meson terms. As we have argued in the
introduction, due to the short range of the exchanged mesons, a zero-range
approximation to their propagators does not make a significant difference in
the results. It also greatly simplifies the numerical calculations.
The Hamiltonian form of the pairing field is, in this case,
\begin{eqnarray}
\bar{\Delta}_t^{\dagger }(\vec{x},\vec{y}) &=& \gamma _0\Delta _t(\vec{x},
\vec{y})\gamma _0  \label{deloc} \\
&=& \delta (\vec{x}-\vec{y}) c_{pair} \left( \frac{g_\sigma ^2}{m_\sigma ^2}
\gamma_0\,\kappa_t(\vec{x})\,\gamma _0
- \left( \frac{g_\omega ^2}{m_\omega ^2}+
\frac{(g_\rho /2)^2}{m_\rho ^2}\right)
\gamma _0\gamma ^\mu \,\kappa_t(\vec{x})\,\gamma _\mu \gamma _0\right)\,,
  \nonumber
\end{eqnarray}
where the anomalous density $\kappa_t(\vec{x})$ is
\begin{equation}
\kappa_t(\vec{x})=\sum_{\varepsilon _{t\gamma }<0,\Omega _{\gamma }>0}\left(
U_{t\gamma }(\vec{x})V_{t\gamma }^{\dagger }(\vec{x}) +BU_{t\gamma }^{*}(%
\vec{x})V_{t\gamma }^T(\vec{x})B^{\dagger }\right) \gamma _0\qquad \qquad
t=p,n.  \label{qmat}
\end{equation}
However, the pairing field, in the given form, does not necessarily
satisfy the antisymmetry condition of Eq.~(\ref{symm}). Applying the
antisymmetry condition to the pairing field, we can reduce it to a
condition on the anomalous density,
\begin{equation}
\kappa_t(\vec x) = B \kappa^T_t(\vec x)B^{\dagger}\,.
\end{equation}
To ensure that this is satisfied, we take the anomalous density to be
\begin{eqnarray}
\kappa_t(\vec x) &\rightarrow& \frac{1}{2}\left(\kappa_t(\vec x) +
B \kappa^T_t(\vec x)B^{\dagger}\right) \qquad\qquad\qquad\qquad\qquad\qquad
t=p,n \label{qsymm}\\
&=& \frac{1}{2}\sum_{\varepsilon _{t\gamma }<0,\Omega _{\gamma }>0}\left(
U_{t\gamma }(\vec{x})V_{t\gamma }^{\dagger }(\vec{x})\gamma_0
+BU_{t\gamma }^{*}(\vec{x})V_{t\gamma }^T(\vec{x})B^{\dagger } \gamma _0
\right.\nonumber \\
&& \qquad\qquad\qquad\left.+\gamma_0 V_{t\gamma }(\vec{x})
U_{t\gamma }^{\dagger }(\vec{x})+\gamma_0 BV_{t\gamma }^{*}(\vec{x})
U_{t\gamma }^T(\vec{x})B^{\dagger }\right)\,.
\nonumber
\end{eqnarray}

We then write the components of the pairing field as
\begin{equation}
\bar{\Delta}_{t}^{\dagger }(\vec{x})=\left(
\begin{array}{cccc}
\delta _{1t}(r_{\bot },z) & 0 & i\,\delta _{3t}(r_{\bot },z) & i\,\delta
_{4t}(r_{\bot },z)e^{-i\varphi } \\
0 & \delta _{1t}(r_{\bot },z) & i\,\delta _{4t}(r_{\bot },z)e^{i\varphi } &
-i\,\delta _{3t}(r_{\bot },z) \\
-i\,\delta _{3t}(r_{\bot },z) & -i\,\delta _{4t}(r_{\bot },z)e^{-i\varphi } &
\delta _{2t}(r_{\bot },z) & 0 \\
-i\,\delta _{4t}(r_{\bot },z)e^{i\varphi } & i\,\delta _{3t}(r_{\bot },z) &
 0 & \delta _{2t}(r_{\bot },z) \label{pard}
\end{array}
\right) ,
\end{equation}
where the four real functions $\delta _{jt}$ are
\begin{eqnarray}
\delta _{1t}(r_{\bot },z) &=&(c_s-c_v)\kappa_{1t}(r_{\bot },z)
-3c_{v}\kappa_{2t}(r_{\bot },z),  \nonumber \\
\delta _{2t}(r_{\bot },z) &=&(c_s-c_v)\kappa_{2t}(r_{\bot },z)
-3 c_{v}\kappa_{1t}(r_{\bot },z),  \label{delcom} \\
\delta _{3t}(r_{\bot },z) &=&(c_{s}+2c_{v})\kappa_{3t}(r_{\bot },z),
\nonumber \\
\delta _{4t}(r_{\bot },z) &=&(c_{s}+2c_{v})\kappa_{4t}(r_{\bot },z),  \nonumber
\end{eqnarray}
with
\begin{equation}
c_{s}=c_{pair}\frac{g_{\sigma }^{2}}{m_{\sigma }^{2}}\qquad \qquad
\mbox{and}\qquad\qquad c_{v}=c_{pair}\left(
\frac{g_{\omega }^{2}}{m_{\omega }^{2}}
+\frac{(g_{\rho }/2)^{2}}{m_{\rho }^{2}}\right).
\end{equation}
The four real components of the anomalous density are
\begin{eqnarray}
\kappa_{1t}(r_{\bot },z) &=&\sum_{\varepsilon _{t\gamma }<0,\Omega _{\gamma
}>0}\left( uf_{t\gamma }^{+}vf_{t\gamma }^{+}
+uf_{t\gamma }^{-}vf_{t\gamma }^{-}\right) ,
\nonumber \\
\kappa_{2t}(r_{\bot },z) &=&\sum_{\varepsilon _{t\gamma }<0,\Omega _{\gamma
}>0}\left( ug_{t\gamma }^{+}vg_{t\gamma }^{+}
+ug_{t\gamma }^{-}vg_{t\gamma }^{-}\right) ,
\label{anoden} \\
\kappa_{3t}(r_{\bot },z) &=&\frac{1}{2}\sum_{\varepsilon _{t\gamma }<0,\Omega
_{\gamma }>0}\left( uf_{t\gamma }^{+}vg_{t\gamma}^{+}
-uf_{t\gamma }^{-}vg_{t\gamma }^{-}+ ug_{t\gamma }^{+}vf_{t\gamma}^{+}
-ug_{t\gamma }^{-}vf_{t\gamma }^{-}\right) ,  \nonumber \\
\kappa_{4t}(r_{\bot },z) &=&\frac{1}{2}\sum_{\varepsilon _{t\gamma }<0,\Omega
_{\gamma }>0}\left( uf_{t\gamma }^{+}vg_{t\gamma }^{-}
+uf_{t\gamma }^{-}vg_{t\gamma }^{+}+ ug_{t\gamma }^{+} vf_{t\gamma}^{-}
+ug_{t\gamma }^{-}vf_{t\gamma }^{+}\right) .  \nonumber
\end{eqnarray}

The Dirac structure of the pairing field is very similar to that
of the $^1$S$_0$ pairing field in symmetric nuclear matter, where
\begin{eqnarray}
\Delta_{nm}(k)  =\gamma_0 \bar{\Delta_{nm}}^{\dagger }(k)\gamma_0 =
\Delta_{S}(k) - \gamma_0\,\Delta_{0}(k)
     - i\gamma_0\vec\gamma\cdot\vec k\,\Delta_{T}(k) \,.
\end{eqnarray}
The upper and lower diagonal components o fthe pairing field, $\delta_{1t}$ and
$\delta_{2t}$ can be directly associated with the linear combinations
$\Delta_S\pm\Delta_0$ of the nuclear matter components, while the 
remaining components, $\delta_{3t}$ and $\delta_{4t}$ are more loosely
related to the contributions of the nuclear matter tensor term, $\Delta_T$.

An overall constant $c_{pair}$ has been introduced in the expression for the
pairing field to compensate for deficiencies
of the interaction parameters and of the numerical calculation. 
The necessity for such a constant is apparent from studies of
pairing in nuclear matter. Nonrelativistic\cite{Bal95,Kho96,Elg96} and
relativistic\cite{Car97} calculations have verified that $^1$S$_0$
pairing in nuclear matter is dominated by the two-nucleon $^1$S$_0$
virtual state. Pairing in nuclear matter is weaker the further the $^1$S$_0$ 
virtual state is from the real axis in the complex-momentum plane. The
location of the virtual state depends on the strength and form of the 
two-nucleon interaction and on the space of states used in the calculation.
In Ref.~\cite{Car97}, various sets of
interaction parameters, even zero-range ones, were shown to furnish
mutually consistent physical values for the pairing gap function,
when they were supplemented with a large momentum cutoff adjusted so as to
place the two-nucleon virtual state at its physical location. 
We expect a condition similar to that in nuclear matter to apply here. However,
as it is extremely difficult to fix the position of the two-nucleon virtual
state within the harmonic oscillator basis that we use, we instead
multiply the pairing field by an overall constant that we expect to be able
to fix independently of the charge and mass of the systems under consideration.
We emphasize that this is not a weakness of our calculations alone, but of any
Hartree-(Fock)-Bogoliubov calculation using a limited space of states and an 
effective interaction, even those using a finite-range one.
The pairing field obtained in such a calculation will depend on both
the interaction and the space of states used and will usually require that
one or the other of these be adjusted on order to obtain reasonable results.
Here, we find it more convenient to introduce an arbitrary constant in the
interaction rather than arbitrarily limit the space of states we use.

With the above simplifications in the self-energy and pairing fields, the
Dirac-Gorkov equations for neutrons and protons reduce to local differential
equations. Their Hamiltonian form is
\begin{equation}
\left(
\begin{array}{cc}
\varepsilon +\mu _{t}-h_{t}(\vec{x}) & \bar{\Delta}_{t}^{\dagger }(\vec{x})
\\
\bar{\Delta}_{t}(\vec{x}) & \varepsilon -\mu _{t}+h_{t}(\vec{x})
\end{array}
\right) \left(
\begin{array}{c}
U_{t}(\vec{x}) \\
\gamma _{0}V_{t}(\vec{x})
\end{array}
\right) =0\,,  \label{hamloc}
\end{equation}
with
\begin{equation}
h_{t}(\vec{x})=-i\vec{\alpha}\cdot \vec{\nabla}+\beta M^*(\vec x) +V_{t}
(\vec{x})\qquad \qquad t=p,n,
\end{equation}
where
\begin{eqnarray}
M^*(\vec x) &=& M-g_{\sigma}\,\sigma(\vec{x})\,,\\
V_t(\vec x) &=& g_{\omega}\,\omega^0(\vec{x}) +
\frac{g_{\rho}}{2}\,2 m_t\,\rho^{00}(\vec{x})
+e\,(1/2+m_t)\,A^0(\vec{x})\,,\nonumber
\end{eqnarray}
and $\bar{\Delta}_{t}^{\dagger}(\vec{x})$ is given in Eq.~(\ref{pard}).

The total energy can now be written in terms of the mean fields as
\begin{eqnarray}
E&=&\int d^{3}x\,\left(\sum_{\varepsilon _{\gamma }<0}
|U_{\gamma }(\vec{x})|^2 (\varepsilon _{\gamma }+\mu )
 -\frac{1}{6}\,g_3\,\sigma(\vec{x})^3
-\frac{1}{4}\,g_4\,\sigma(\vec{x})^4
+\frac{1}{2} g_{\sigma}\,\sigma(\vec{x})\,\rho_s(\vec{x})\right.
 \label{eneqloc}\\
& & \left.\qquad
-\frac{1}{2}g_{\omega}\,\omega^0(\vec{x})\,\rho_B(\vec{x})
-\frac{1}{2}\frac{g_{\rho}}{2}\,\rho^{00}(\vec{x})\,\rho_3(\vec{x})
-\frac{1}{2}e\,A^0(\vec{x})\,\rho_c(\vec{x})
+\frac{1}{2}\sum_{t}\mbox{Tr}[\bar{\Delta}_t^{\dagger}(\vec{x})\,
\kappa_{t}(\vec{x})]\right)
\nonumber\\
& & \qquad\qquad-E_{cm}\,.\nonumber
\end{eqnarray}
In the expression above, we have also subtracted the harmonic oscillator
estimate to the center-of-mass motion,
\begin{equation}
E_{cm}=\frac{3}{4}\hbar\omega_0 =\frac{3}{4}\frac{41}{A^{1/3}}\,\,\mbox{MeV},
\end{equation}
in order to obtain an expression for the total internal energy of the nucleus.

\section{Numerical solution of the DHB equation}

We solve the Dirac-Gorkov and the Klein-Gordon equations by expanding the
fields as well as the wavefunctions in complete sets of eigenfunctions of
harmonic oscillator potentials. In actual calculations, the expansion is
truncated at a finite number of major shells, with the quantum number of
the last included shell denoted by $N_F$ in the case of the fermions and
by $N_B$ for the bosons. The maximum values are selected so as to assure
the physical significance of the results obtained. The same procedure
has been used by many researchers, among them, by Vautherin\cite{Vau80}
in the nonrelativistic Hartree-Fock approximation, by Ghambir et
al.\cite{Gam90} in the relativistic mean field + BCS approach and by
Lalazissis et al.\cite{Lal98c,Vre98b,Lal98d} in the RHB approach.

The spinors of the Dirac-Gorkov equation are expanded in terms of the
eigenfunctions of an axially-deformed harmonic-oscillator potential,
\begin{equation}
V_{osc}(r_{\bot},z)=\frac 12M\omega _z^2z^2+\frac 12M\omega _{\bot }^2r^2.
\end{equation}
The oscillator constants are taken as
\begin{equation}
\beta _z=\frac 1{b_z}=\sqrt{\frac{M\omega _z}\hbar }\qquad \beta _{\bot }=%
\frac 1{b_{\bot }}=\sqrt{\frac{M\omega _{\bot }}\hbar }.
\end{equation}
with volume conservation relating the two constants to that of a
spherically-symmetric potential $b_{\bot }^2b_z=b_0^3$.

The eigenfunctions of the deformed harmonic oscillator can be written
explicitly as,
\begin{equation}
\Phi_{\alpha} (\vec{r})=\psi _{n_r}^{m_l}(r_{\bot})\,\psi _{n_z}(z)\, \frac{%
e^{i\,m_l\,\varphi }}{\sqrt{2\,\pi }}\chi _{m_s}\chi _{m_t}
\end{equation}
where $\alpha$ denotes the complete set of quantum numbers ($n_r$, $m_l$, $%
n_z$, $m_s$, and $m_t$) and
\begin{eqnarray}
\psi _{n_r}^{m_l}(r_{\bot}) &=&\frac{N_{n_r}^{m_l}}{b_{\bot }}\sqrt{2}\,\eta
^{m_l/2}\,L_{n_r}^{m_l}(\eta )\,e^{-\eta /2}\qquad \mbox{with}\qquad
\eta=\left(\frac{r}{b_{\bot}}\right)^2  \label{func} \\
\psi _{n_z}(z) &=&\frac{N_{n_z}}{\sqrt{b_z}}H_{n_z}(\xi )e^{-\xi ^2/2}\qquad %
\mbox{with}\qquad \xi=\frac{z}{b_z}  \nonumber
\end{eqnarray}
In Eq.~(\ref{func}), $L_{n_r}^{m_l}(\eta )$ and $H_{n_z}(\xi )$ are Hermite
and associate Laguerre polynomials\cite{Abr70}, with the normalization
constants, $N_{n_r}^{m_l}$ and $N_{n_z}$, given in Ref. \onlinecite{Gam90}.
In these equations, $n_r$ and $n_z$ are the number of nodes in the $r$ and $%
z $ directions, and $m_l$ and $m_s$ are the projections of angular momentum
and spin on the $z$ axis. The third component of the total angular momentum $%
\Omega _{\gamma }$ and the parity $\pi $ are then defined as,
\begin{equation}
\Omega _{\gamma }=m_l+m_s\qquad \pi =(-1)^{n_z+m_l}.
\end{equation}

We expand the Pauli components of the Dirac spinors, $uf_{t\gamma
}(r_{\bot},z)$, $ug_{t\gamma}(r_{\bot},z)$, $vf_{t\gamma }(r_{\bot},z)$, and
$vg_{t\gamma}(r_{\bot},z)$, in terms of the oscillator eigenfunctions.
Inserting these expansions into the Dirac-Gorkov equation
Eq.~(\ref{hamloc}), we can reduce the equation to the diagonalization problem
of a symmetric matrix and calculate the Hartree densities of Eq.~(\ref{rhoss})
and components of the anomalous density of Eq.~(\ref{anoden}) 
The fields of the massive mesons are expanded in a manner similar to the
fermion expansion, with the same deformation parameter $\beta_{0}$ but
a smaller oscillator length of $b_{B}=b_{0}/\sqrt{2}$. 
The Coulomb field is calculated directly in configuration space.
In short, the method used is a direct generalization of that described in
Ref.\onlinecite{Gam90}, where more details may be found.

\section{Numerical Results}

The parameters required to perform numerical calculations are the
nucleon and meson masses, the meson-nucleon coupling constants and
the factor $c_{pair}$ that multiplies the pairing interaction. The
calculations that we present were performed using the masses
and coupling constants of the NL3 potential\cite{Lal97}. We performed
calculations for several values of the factor $c_{pair}$. These
permit us to study the extent to which physical
observables depend on the pairing interaction and to choose the value
of the parameter $c_{pair}$ that best fits the observables.

As the mesons fields and the nucleon wavefunctions are expanded in a
deformed basis of harmonic oscillator states, we must also specify the
number of major oscillator shells to be used in the expansions for
fermions, $N_{F}$, and bosons, $N_{B}$, as well as the basis deformation
parameter, $\beta_{0}$. Here, we work, for the most part, in a basis of
12 major oscillator
shells for fermions and 24 for bosons. We use the standard expression
for the oscillator frequencies, $\hbar\omega_{0}=41A^{-1/3}$ MeV, and
spherical bases, with $\beta_0=0$, in all calculations. Although the
number of major oscillator shells is not as large as might be
desired, it does seem to be large enough to obtain reasonable values
for the observables studied.

In order to analyze the characteristics of of the single-particle
levels, we define several related average values. First, in deference
to the standard nonrelativistic notation, we define the occupation
probability of each of the two states of the level with frequency
$\varepsilon_{t\gamma}<0$ as

\begin{equation}
v_{t\gamma }^{2}=\int d^{3}x\,|U_{t\gamma }(\vec{x})|^{2},
\end{equation}
so that we also have

\begin{equation}
u_{t\gamma }^{2}=1-v_{t\gamma }^{2}=\int d^{3}x\,|V_{t\gamma }(\vec{x})|^{2}.
\end{equation}
We define the energy of a single-particle level as

\begin{equation}
E_{t\gamma }=\int d^{3}x\,\left( U_{t\gamma }^{\dagger }(\vec{x}%
)\,h_{t}U_{t\gamma }(\vec{x})+V_{t\gamma }^{\dagger }(\vec{x}%
)\,h_{t}V_{t\gamma }(\vec{x})\right) ,
\end{equation}
in which we take advantage of the normalization of the state vector
to avoid normalizing the result. This is not possible for the pairing
term, for which we define the gap parameter of a single-particle
level as
\begin{equation}
\Delta _{t\gamma }=-\int d^{3}x\,U_{t\gamma }^{\dagger }(\vec{x})\,\bar{%
\Delta}_{t}^{\dagger }\gamma _{0}V_{t\gamma }(\vec{x})\,/\,u_{t\gamma
}\,v_{t\gamma }.
\end{equation}
The definition of the gap parameter is fragile and is subject to
numerical error when  $v_{t\gamma }\rightarrow 1$ or
$u_{t\gamma }\rightarrow 1$. We define an average gap parameter
for the neutrons and protons of a nucleus as
\begin{equation}
<\Delta _t>=\sum_{\varepsilon_{\gamma}<0}\,\Delta _{t\gamma } \,v^2_{t\gamma}
\left/ \sum_{\varepsilon_{\gamma}<0}\,v^2_{t\gamma}\right..
\end{equation}
We note that, unlike in the BCS approximation, no relation exists here
between the occupation probabilities, $v_{t\gamma }^{2}$ and
$u_{t\gamma }^{2}$,
the energy $E_{t\gamma }$, and the gap parameter $\Delta _{t\gamma }$.
However, these averages still furnish a good description of the most
important characteristics of the single-particle levels.

In the following, we first analyze pairing in the isotopes of
tin and then discuss similar results for the isotopes of nickel
and calcium, all of which are spherical. The isotopes of tin
and nickel have already been the object of two thorough studies
\cite{Lal98c,Dob96}. We then analyze
deformation and pairing of the $N$=28 isotonic chain, which has
been the object of two recent studies\cite{Lal98d,Wer94,Wer96}.
Finally, we turn our attention to nuclei in the region of the $Z\approx 40$
subshell closure, the isotopes of Kr and Sr, in particular. These
have been the object of many previous
studies.\cite{Hir93,She93,Lal95,Bon85,Bon91}

\subsection{Spherical Nuclei}

Pairing in the Sn and Ni isotopes has been studied extensively, due to the
simplification provided by their spherical symmetry. In Ref. 
\onlinecite{Dob96}, Dobaczewski et al. describe a very complete study of 
pairing in the Sn isotopes, based on a nonrelativistic Hartree-Fock-Bogoliubov
approximation, in which they examine the differences in pairing due to the use
of various effective interactions and due to coupling to the particle
continuum, as well as the effects of these on experimental observables.
In Ref. \onlinecite{Lal98c}, Lalazissis et al. compare RHB calculations of both
the odd and even isotopes in the Sn and Ni chains with the experimental data,
obtaining quite good agreement. In another study\cite{Men98}, Meng compared
RHB calculations using the finite-range Gogny D1S and a zero-range interaction
and found good agreement between the two and with the experimental data.

We begin our study of spherical nuclei with the tin isotopes. 
We performed calculations of the ground states of the even isotopes from
$^{100}$Sn to $^{176}$Sn, that is, from the closed neutron shell at
$N$=50 to the closed shell at $N$=126. We present calculations for
three values of the parameter $c_{pair}$. The objective of
this study was to identify the observables sensitive to the parameter
$c_{pair}$ (and, thus, to the pairing) and to adjust the parameter accordingly.

\begin{figure}[htbp]
\begin{center}
\epsfig{file=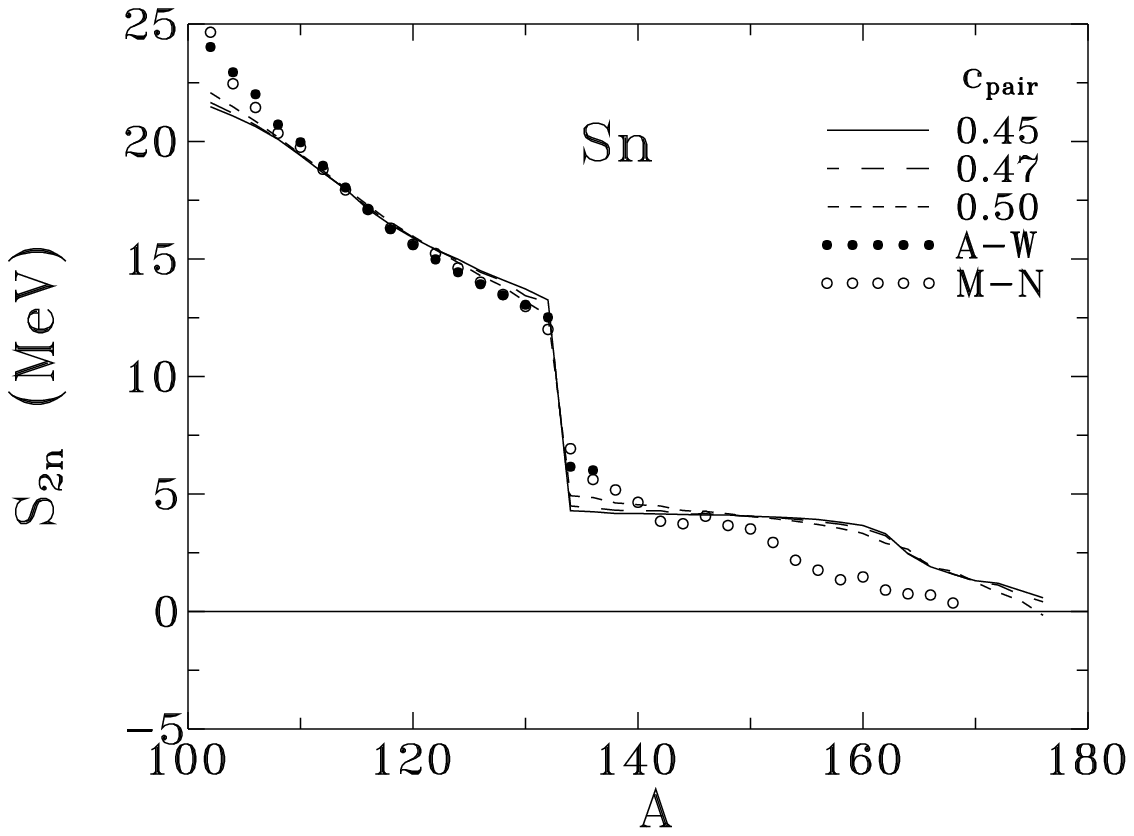,height=7.0cm}
\caption{ Two-neutron separation energy of the even isotopes of Sn,
as a function of the mass number $A$, for three values of the parameter
$c_{pair}$. The values obtained from the compilation
of experimental masses of Audi and Wapstra\protect\cite{Aud95} (solid circles)
and the M\"oller-Nix systematics\protect\cite{Mol95} (open circles)
are also shown.
\label{fig1}}
\end{center}
\end{figure}

We present in Fig.~\ref{fig1} the two-neutron separation energy of
the even Sn isotopes in the mass range from $A$=100 to $A$=176, calculated
for the values $c_{pair}$=0.45, 0.47 and 0.50. We compare our
calculations with the two-neutron separation energies obtained using the
ground state masses tables of M\"oller-Nix\cite{Mol95} and of 
Audi-Wapstra\cite{Aud95}. The M\"oller-Nix values for ground state masses were
calculated using an extended finite-range droplet model with parameters
adjusted to the experimental ground state masses. The Audi-Wasptra values for
the ground state masses are essentially the experimental ones, with some 
extension to proton- and neutron-rich nuclei based on systematics.  
We verify that all three calculations follow the trend of the M\"oller-Nix
and Audi-Wapstra values. The
two-neutron separation energy is seen to be almost independent of the
strength of the pairing field, with the calculation using $c_{pair}$=0.50
providing only slightly better agreement with the data than the others. All
the calculations underestimate $S_{2n}$ in the region of $A$=100. This
discrepancy might be attributable to a deficiency in the isospin dependence
of the NL3 parameter
set. It could also be due to effects that are not included in the
calculation, such as neutron-proton correlations, which are suspected of being
of importance in $N \approx Z$ nuclei. Our two-neutron separation energies are 
in good agreement with the RHB ones of Ref. \onlinecite{Lal98c}. However, 
such consistency is to be expected, given the relative insensitivity of the
two-neutron separation energy to the pairing, since both calculations use the
NL3 parameter set.

The calculated two-neutron separation energies remain almost constant from
the shell closure at $A$=132 to about $A$=160, in contrast to the
M\"oller-Nix values, which decrease slowly. The SIII and Sk$\delta$ Skyrme 
interaction calculations of Ref. \onlinecite{Dob96} also furnish a
relatively constant separation energy above the $A$=132 shell closure,
which extends to even higher values of the mass, although most of their
calculations are in agreement with the M\"oller-Nix systematics. 
The nucleus $^{176}$Sn is unbound in the calculation with $c_{pair}$=0.50, as
it is in all of the calculations of Ref. \onlinecite{Dob96}.  In our 
calculations, this is due to the fact that $^{174}$Sn is more tightly bound 
for $c_{pair}$=0.50 than for the other values of
the parameter. The binding energy of the magic nucleus $^{176}$Sn is the same
for all values of the parameter $c_{pair}$.

\begin{figure}[htbp]
\begin{center}
\epsfig{file=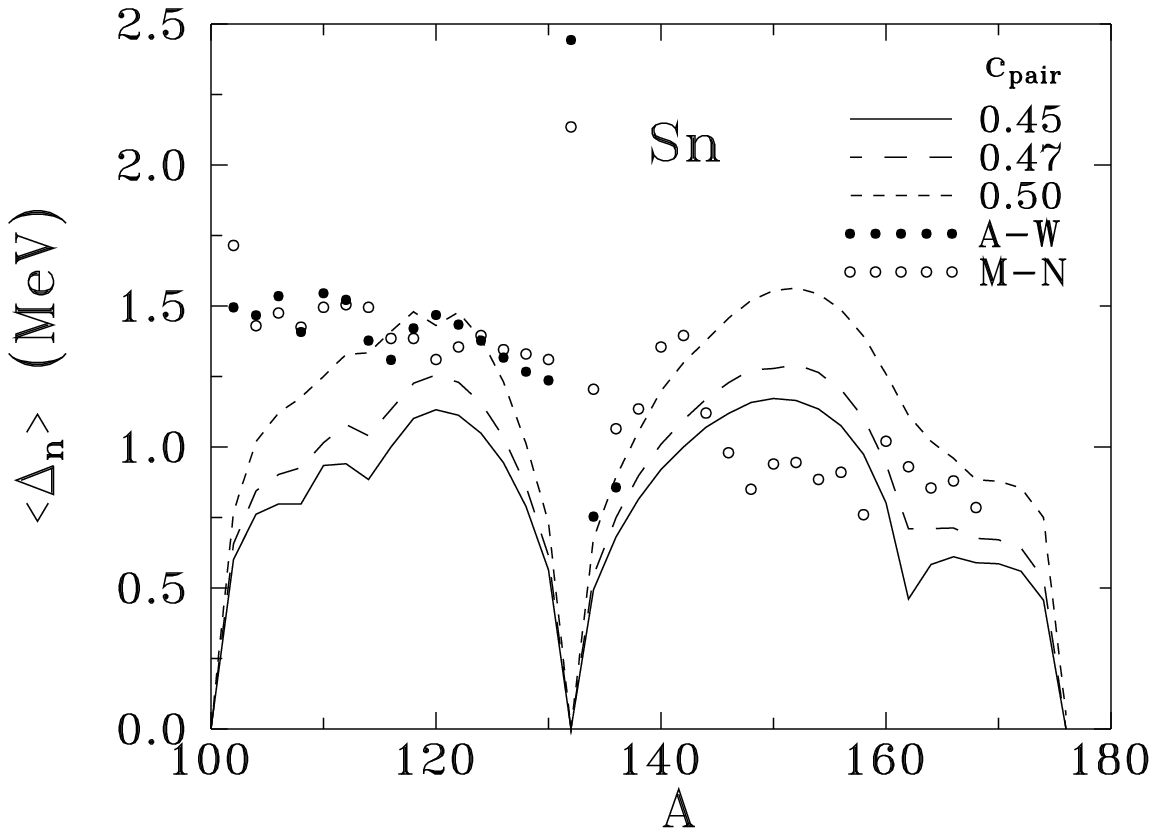,height=7.0cm}
\caption{ Mean value of the neutron gap parameter of the even
isotopes of Sn, as a function of the mass number $A$, for three
values of the parameter $c_{pair}$ (lines). The odd-even mass
differences obtained from the compilation of experimental masses
of Audi and Wapstra\protect\cite{Aud95} (solid circles) and from
the M\"oller-Nix systematics\protect\cite{Mol95} (open circles) are
also plotted.
\label{fig2}}
\end{center}
\end{figure}

In Fig.~\ref{fig2} we show the average value of the neutron gap parameter of
the even
Sn isotopes as a function of the mass number $A$, for the same three values of
$c_{pair}$. The gap parameter possesses a clear dependence on the value of
$c_{pair}$. We also show in the figure the M\"oller-Nix and experimental 
Audi-Wapstra values for the standard
estimate of the neutron gap parameter as the difference between the binding
energies of an even-even nucleus and its odd mass neighbors,
\[
<\Delta_n(Z,A)>=B(Z,A)-\frac{1}{2}(B(Z,A-1)+B(Z,A+1))
\]
where $B(Z,A)$ is the binding energy. The calculation with $c_{pair}$=0.50
shows reasonable agreement with the M\"oller-Nix values in the region of
the $N$=82, $A$=132, shell closure, but tends to underestimate the 
M\"oller-Nix values well below the shell closure and overestimate these
values well above the shell closure. 

The neutron shell closures at $N$=50,82, and 126 are clearly visible in
Fig.~\ref{fig2}. At each of values of the neutron number,
calculated neutron gap parameter goes to zero. We note that the experimental
gap parameter has maximum, rather than a minimum, at each shell closure, 
reflecting the local maximum of the binding energy that occurs there.
Several subshell closures are
also visible in the calculations, at values of the mass at which the gap
parameter reaches a non-zero minimum. These occur at $N$=58, $A$=108,
between the $1g7/2$ and $2d5/2$ levels, at $N$=64, $A$=114, between the
$2d5/2$ level and the remaining levels of the $4\hbar\omega$ shell, and at
$N$=112, $A$=162, between the $5\hbar\omega$ shell and the $1i13/2$ level.
Subshell closures are more visible when the pairing interaction is weaker
and, thus, more sensitive to the energy differences between the levels.

Average values of the neutron gap parameter of the even Sn isotopes
obtained in RHB
and nonrelativistic calculations were presented in Refs. \cite{Lal98c}
and \cite{Dob96}, respectively. The calculations using the Gogny interaction,
both the RHB and the nonrelativistic ones, furnish a large
value of the gap parameter -- above 2 MeV -- and show no subshell structure.
Were we to increase the strength of the pairing interaction so as to obtain
similar values for the pairing gap, our calculations would also show no shell
substructure in the average pairing gap. The nonrelativistic calculations of
Ref. \onlinecite{Dob96}, using Skyrme pairing interactions produced results
in closer agreement with ours. These calculations also show structure due
to subshell closures, although the structure is different from that seen here,
possibly due to differences in the spin-orbit splitting of the levels, due to
the different mean fields.

\begin{figure}[htbp]
\begin{center}
\epsfig{file=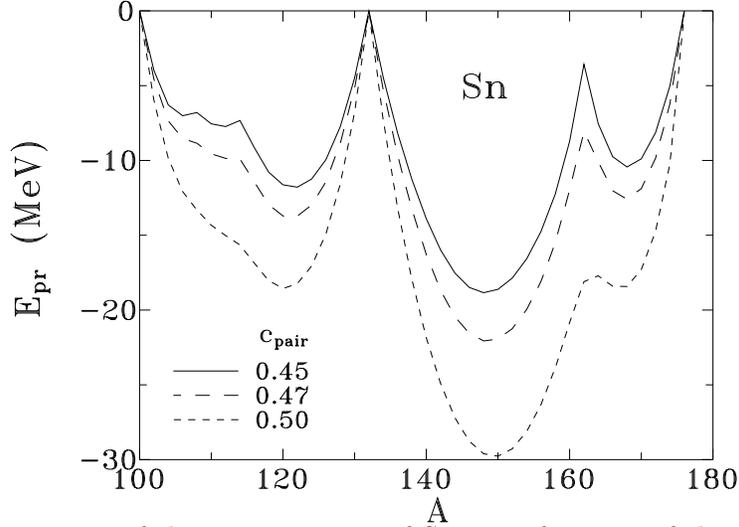,height=7.0cm}
\caption{ Pairing energy of the even isotopes of Sn, as a function of
the mass number $A$, for three values of the parameter $c_{pair}$.
\label{fig3}}
\end{center}
\end{figure}

In Fig.~\ref{fig3}, we show the pairing energy of the even Sn isotopes as a
function
of the mass number, obtained using the same three values of $c_{pair}$. The
pairing energy is defined as
\[
E_{pr}=-\frac{1}{2}\sum_{t}\int dx^3\,
\mbox{Tr}[\bar{\Delta}_{t}^{\dagger }(\vec{x})\,\kappa _{t}(\vec{x})],
\]
where  $\kappa_t$ and $\bar{\Delta}_{t}^{\dagger }$ are given in
Eqs. (\ref{qsymm}) and (\ref{pard}), respectively. We observe that the pairing
energy $E_{pr}$ displays a dependence on the parameter $c_{pair}$ and a shell
structure similar to those seen in Fig.~\ref{fig2} for the average gap
parameter. However, $E_{pr}$ is observable only through its effects
on the binding energy, which diminishes its utility as a means of determining
the parameter $c_{pair}$.

\begin{figure}[htbp]
\begin{center}
\epsfig{file=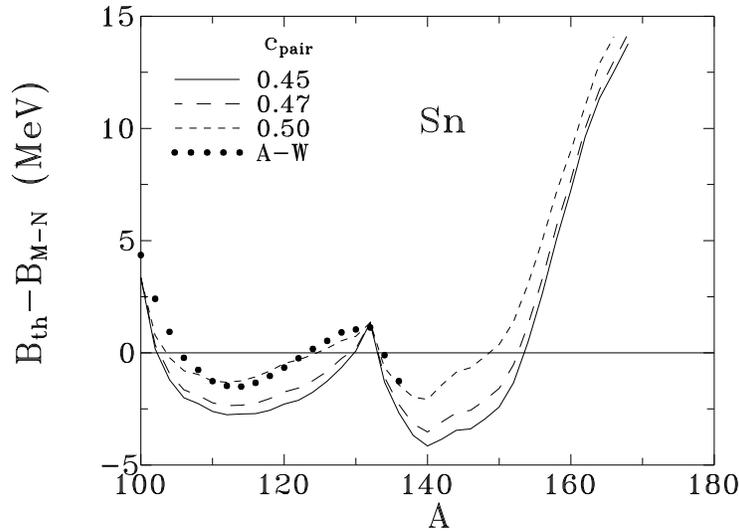,height=7.0cm}
\caption{  Difference between the calculated binding energy and the
M\"oller-Nix binding energy\protect\cite{Mol95} for
the even isotopes of Sn, as a function of the mass number $A$, for three
values of the parameter $c_{pair}$ (lines). The difference between the
calculated binding energy and the Audi-Wapstra experimental binding energy%
\protect\cite{Aud95}, for $c_{pair}$=0.500, is also plotted (circles).
\label{fig4}}
\end{center}
\end{figure}

In Fig.~\ref{fig4} we present the difference between the calculated value of
the binding energy of the even Sn isoptopes and the M\"oller-Nix value, as a
function of the mass, using the same three values of $c_{pair}$ used before.
We also present the difference between the calculated value for
$c_{pair}$=0.50 and the Audi-Wapstra values, where the latter exist.
In the mass range of the valley of stability, the difference between the two
binding energies can be minimized, to a certain point, through an appropriate
choice of $c_{pair}$. For the tin isotopes,
a value of $c_{pair}$ close to 0.50 seems to yield the best average 
agreement. A value much larger than this would destroy the good agreement
obtained for the gap parameter of Fig.~\ref{fig2}. However, a larger value
of $c_{pair}$ would smooth out the two dips in the binding energy difference
that roughly follow the mass dependence of both the average gap parameter
and the pairing energy. This would yield pairing gaps in better agreement with
the Gogny D1S ones of Ref. \onlinecite{Lal98c}. We then might expect to be
able to reduce the remaining discrepancy by adjusting the mean field
parameters.

In Fig.~\ref{fig4}, the calculated value of the binding energy
is larger than the experimental/systematic values in the $N\approx Z$
region and reaches a difference of about 4 MeV for the magic nucleus
$^{100}$Sn, a result which is independent of the value of $c_{pair}$.
The sharp increase of the binding-energy difference for $N\approx Z$
is consistent with the discrepancies
between the calculated and the M\"oller-Nix values of the two-neutron
separation energy, shown in Fig.~\ref{fig1}. We can attribute the
discrepancies to deficiencies in the isospin dependence of the NL3 parameters.
If we were to include neutron-proton pairing in the calculations, the
difference would only increase as the pairing would bind even more the
already overbound $N\approx50$ nuclei.
The large differences in binding energy on the neutron-heavy side
of the curve could be due as much to deficiencies in the 
M\"oller-Nix systematics as in the NL3 parameters.

\begin{figure}[htbp]
\begin{center}
\epsfig{file=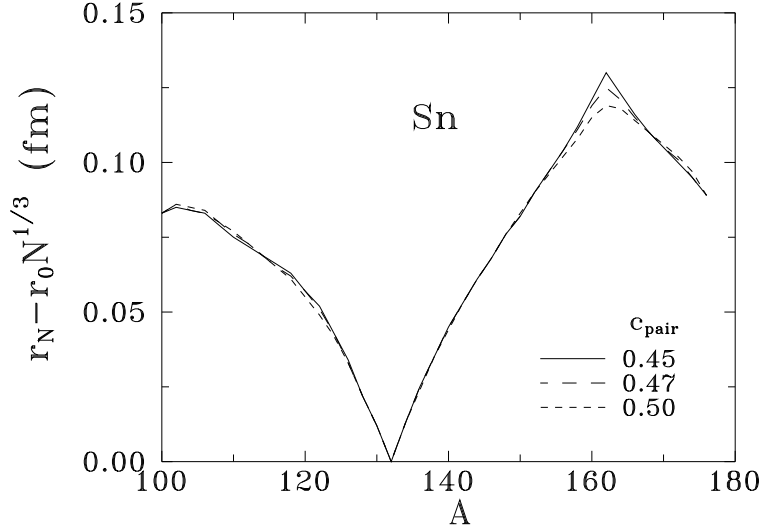,height=7.0cm}
\caption{ Deviation of the root mean square neutron radius from the
systematic value $r_0N^{1/3}$, for the even isotopes of Sn, as a
function of the mass number $A$, for three values of the parameter
$c_{pair}$. The reduced radius, $r_0$, was adjusted to the root mean
square neutron radius of $^{132}$Sn.
\label{fig5}}
\end{center}
\end{figure}

In Fig.~\ref{fig5} we present the deviation from the systematic value,
$r_{n0}N^{1/3}$, of the root mean square neutron radius of the even isotopes
of Sn, as a function of the mass. The deviation displays
a clear minimum at the principal shell closure at $N$=82. Although the
deviation also tends to decrease near the shell closures at $N$=50 and
$N$=126, the effect of these more extreme shell closures is much smaller than
that of the $N$=82 one, within the stability valley. The deviation
from the systematic value also possesses a sharp maximum at
$N$=112, between the $5\hbar\omega$ shell and the $1i13/2$ level. We recall
that the average gap parameter and the pairing energy display a minimum at the
same subshell closure. Other subshell closures also appear in the deviation
of the radius, but only as changes in its slope. The deviation from the
systematic value of the root mean square mass radius of the even Sn isotopes
shows the same structure, but to a lesser degree, and is not shown here. The
deviation from the systematic of the root mean square proton radius, which is
also not shown, increases monotonically with the mass number.

We find that the deviations of the root mean
square radii are almost independent of the parameter $c_{pair}$ and, thus,
of the pairing interaction. Slightly larger variations have been seen in a
comparison between RHB and Hartee+BCS calculations\cite{Lal98d}. Still, with
the exception of a few very special nuclei, such as $^{11}$Li, the root
mean square radii of a nucleus would seem to be determined almost
exclusively by its mean field.

As the second example of a spherical isotopic chain, we studied the even
isotopes of nickel, performing calculations from $^{48}$Ni, at the $N$=20
shell closure, to $^{100}$Ni, two neutrons beyond the subshell closure at
$N$=70. We again performed calculations for three values of the parameter
$c_{pair}$, their values in this case being $c_{pair}$=0.49, 0.50, and
0.52. Here we wanted to verify the generality of the observations made in the
case of the Sn isotopes.

The calculations of the two-neutron separation energy for the Ni isotopes
describe the tendency of the data fairly well,
reproducing the discontinuities in separation energy that occur at the shell
closures at $N$=28 and $N$=50. The calculated values of the
(two-neutron) separation energy are almost independent of the parameter
$c_{pair}$ and underestimate the mass dependence of the separation energy
at very low and very high values of the neutron excess, as in the case of Sn.
Not surprisingly, our values are in good agreement with the RHB ones of 
Ref.~\onlinecite{Lal98c}, which also use the NL3 parameter set. They also
agree well with the results of 
Ref.~\onlinecite{Men98}, obtained with the NLSH parameter set\cite{Sha93},
using Gogny D1S and density-dependent zero-range pairing interactions. 

\begin{figure}[htbp]
\begin{center}
\epsfig{file=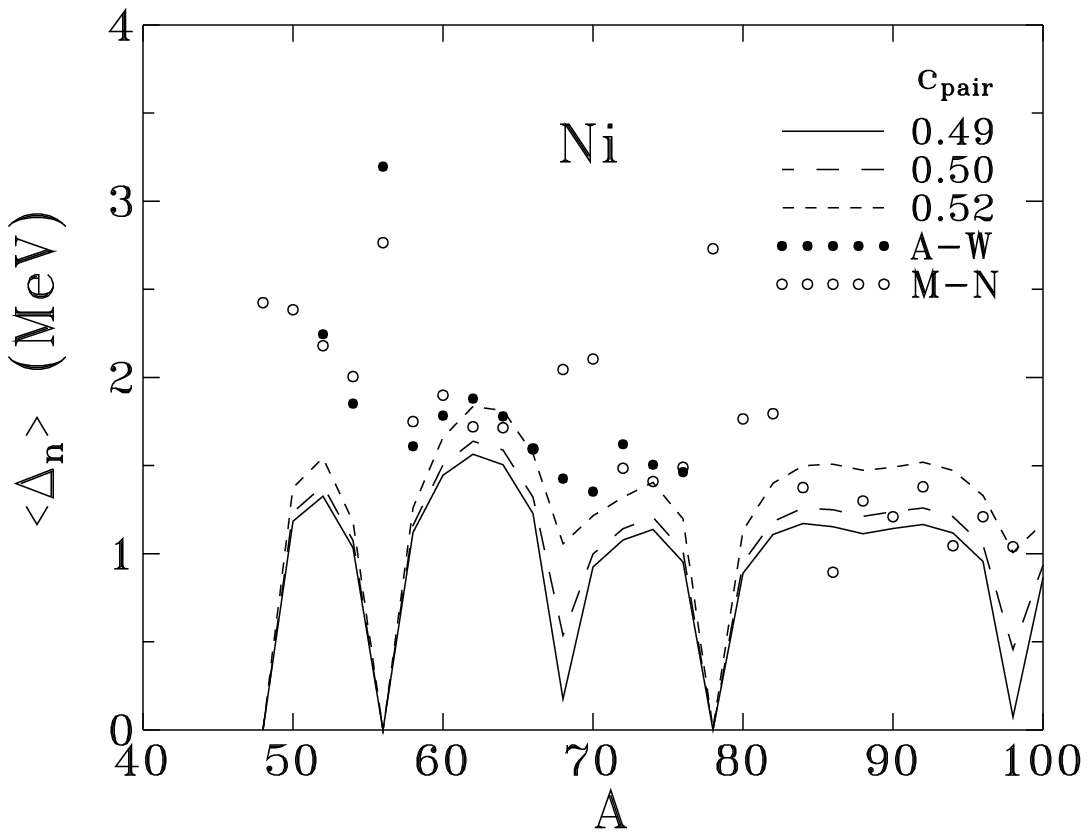,height=7.0cm}
\caption{ Mean value of the neutron gap parameter of the even
isotopes of Ni, as a function of the mass number $A$, for three
values of the parameter $c_{pair}$ (lines).  The odd-even mass
differences obtained from the compilation of experimental masses
of Audi and Wapstra\protect\cite{Aud95} (solid circles) and from
the M\"oller-Nix systematics\protect\cite{Mol95} (open circles) are
also plotted.
\label{fig7}}
\end{center}
\end{figure}

In Fig.~\ref{fig7}, we present the mean value of the neutron gap parameter of
the even
Ni isotopes as a function of the mass, for the same three values of $c_{pair}$.
Again, we find a clear dependence of the gap parameter on the value of
$c_{pair}$. The calculation with $c_{pair}$=0.52 shows the best agreement
with the M\"oller-Nix and Audi-Wapstra
values in the region above the shell closure at $N$=28, $A$=56. All of the
calculations are in disagreement with the experimental/systematic values in
the region of the magic number $N$=28. This again suggests that shotr-range
neutron-proton correlations may be important when $N\approx Z$.

In Fig.~\ref{fig7}, we note that the neutron gap parameter is zero at
the neutron
shell closures at $N$=20, 28, and 50. The subshell closures are more marked
here than in the case of tin. For Ni, these appear at $N$=40, $A$=68, between
the $3\hbar\omega$ shell and the $1g9/2$ level and at $N$=70, $A$=98, between
the $4\hbar\omega$ shell and the $1i11/2$ level. As occurred in the
calculations of the isotopes of Sn, the value of the average gap parameter at
the subshell closures increases with $c_{pair}$.

Values of the average neutron gap parameter of the Ni isotopes between
$N$=28 and $N$=50,
obtained using the Gogny interaction in a RHB calculation, were presented
in Ref.\onlinecite{Lal98c}. As in the case of Sn, the RHB calculation furnishes
values of the gap parameter greater than 2.5 MeV, which are quite large.
The calculation presents a subshell structure similar to that with
$c_{pair}$=0.52 in Fig.~\ref{fig7}, but almost 50\% larger in magnitude.

As in the case of the Sn isotopes, the pairing energy of the even Ni isotopes
reflects the shell structure and the dependence on $c_{pair}$ observed
in the average gap parameter, but furnishes no new information. The root mean
square radii reflect the same shell structure, although with less clarity,
and are almost independent of the value of $c_{pair}$.

We also examined the Ca isotope chain, performing
calculations of the even isotopes of Ca from $N$=8, $A$=28, to $N$=50, $A$=70,
for two values of the parameter $c_{pair}$. Comparisons between the two
calculations and the Audi-Wapstra and M\"oller-Nix systematics resulted in
conclusions similar to those obtained for the tin and nickel isotopes. In
particular, the two-neutron separation energies were well described in the
two calculations, which show almost no dependence on the parameter $c_{pair}$.
A value of $c_{pair}$=0.55 best agrees with the gap parameters obtained from
the Audi-Wapstra systematics, which, in this case, lie about 20\% below the
M\"oller-Nix values.

In summary, we found our calculations of the root mean square radii and the
two-neutron separation
energies of spherical nuclei to be relatively independent of the strength
of the pairing interaction. As the experimental data for these quantities
are well fit by the NL3 parameter set used here, it is not 
surprising that our calculations describe them well. 
We observed that the pairing energy is only
observable through its effects on the binding energy and that our calculations
are not as successful at describing the latter as they are with the radii and
the separation energies. A reasonable fit to the binding energies will depend
on the adequate choice of the parameter $c_{pair}$
as well as further adjustments in other interaction
parameters. We found that the binding energy and the mean value of the gap
parameter, to a certain point, are the observables that can be used as
guides to a study of pairing in spherical nuclei.

\subsection{Several deformed light nuclei}

In this section, we analyze pairing and deformation in the $N=28$ isotonic
chain. The measured quadrupole deformations for these nuclei characterize
$^{48}$Ca as spherical but $^{46}$Ar and $^{44}$S as deformed, implying
suppression of the $N=28$ magic number. The
deformation can be explained, in this case, by the close proximity of
the $1f7/2$ level to the $2p$ levels in the $4\hbar\omega$ shell. In the
nucleus $^{48}$Ca, the spherical proton configuration constrains the
neutron configuration to also remain spherical. In the $^{46}$Ar and
$^{44}$S nuclei, the partially filled proton shell perturbs the neutrons
sufficiently for them to prefer a deformed
configuration in which the $1f7/2$ and $2p$ orbitals are partially occupied.

Recent theoretical studies describe the experimental data for these nuclei
reasonably
well with nonrelativistic and relativistic mean field + BCS\cite{Wer94,Wer96}
calculations and with RHB calculations using the Gogny potential\cite{Lal98d}.
In Refs. \onlinecite{Wer94} and \onlinecite{Wer96}, Werner et al. use both a
Skyrme-Hartree-Fock approach and a relativistic mean field model, with the
NLSH parameter set, to study various istopic chains in the $N\approx 28$
region. They include pairing as a small (75 keV) constant pairing gap in a 
BCS formalism. They found the isotones $^{42}$Si, $^{44}$S and $^{46}$Ar to 
show evidence of shape coexistence in both approaches, although the
deformation at the energy minimum and its energy difference with the excited
minimum varied
between the two calculations. Lalazissis et al. used the RHB approach to
perform a similar study of the $N\approx 28$ isotopes in Ref. 
\onlinecite{Lal98d}. Using the NL3 parameter set, they studied the even 
isotopes of Mg through Cr. They observed prolate-oblate staggering in the 
$N=28$ isotones below $^{48}$Ca, similar to that seen in the RMF calculations
of Werner et al. and similar signs of shape coexistence: the binding energies
versus quadrupole deformation of the isotones display two minima or a
very flat single minimum.

\begin{figure}[htbp]
\begin{center}
\epsfig{file=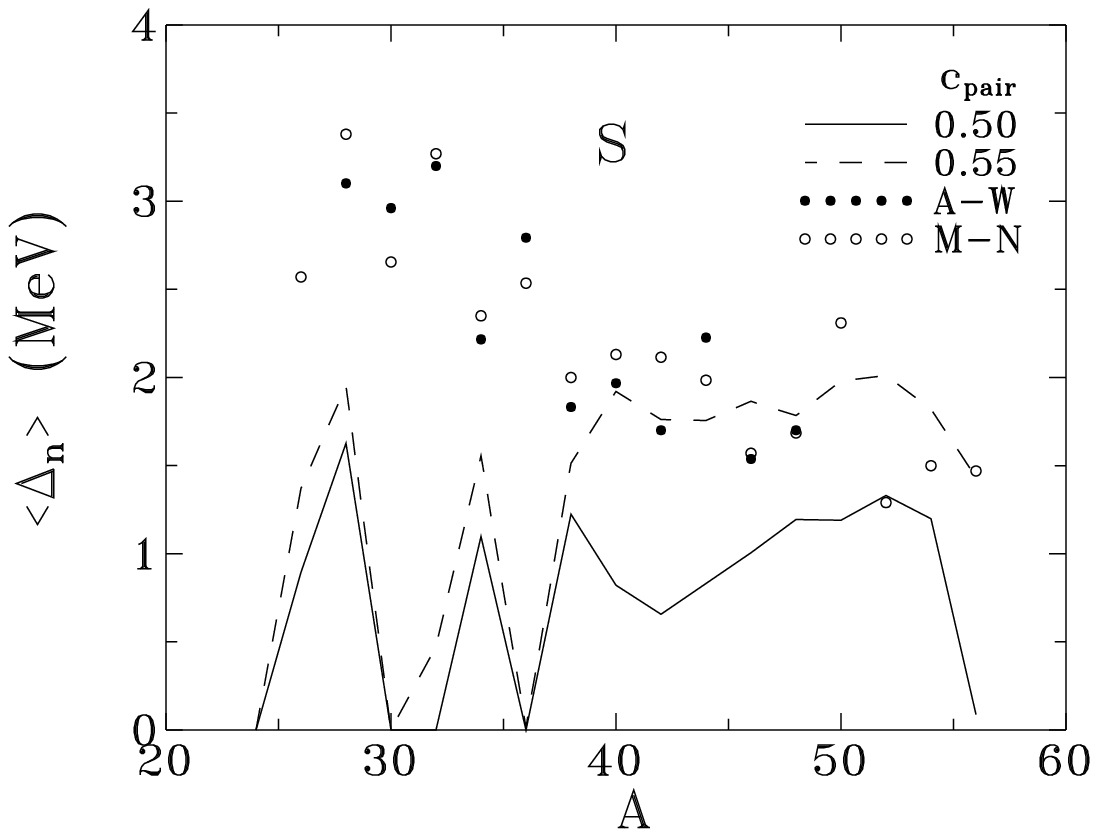,height=7.0cm}
\caption{ Average value of the gap parameter of the even isotopes of
S, as a function of the mass number $A$, for two values of the
parameter $c_{pair}$ (lines). The odd-even mass
differences obtained from the compilation of experimental masses
of Audi and Wapstra\protect\cite{Aud95} (solid circles) and from
the M\"oller-Nix systematics\protect\cite{Mol95} (open circles) are
also plotted.
\label{fig12}}
\end{center}
\end{figure}

We present here the average values of the gap parameter of the
even-even S isotopes, in Fig.~\ref{fig12} as a
function of the mass, for two values of the parameter $c_{pair}$.
We observe that both calculations fall well below the M\"oller-Nix
and Audi-Wapstra values at lower values of the mass. However, the
isotopes that are of interest at the moment are in the region of large 
neutron excess, $A$$\gtrsim$40, where the average values of the
calculated gap parameter and the experimental/systematic values are in
reasonable agreement. In particular, we see in Fig.~\ref{fig12}, that
the M\"oller-Nix and Audi-Wapstra values of the gap in $^{44}$S fall close
to the calculation with $c_{pair}$=0.55. The average value of the gap
parameters of the other isotopes that we will examine show the same
reasonable agreement with the experimental/systematic values in the mass
range of interest.

To study in detail the dependence of the binding energy on the
quadrupole deformation, we perform calculations at several fixed
values of the deformation. To do this, we include in the Lagrangian a
term that is quadratic in the difference between the quadrupole moment of the
nucleus and the desired value of the moment. The
solution of the equation of motion including this additional potential
term tends to take a value which minimizes its contribution,
thereby yielding a quadrupole moment close to the desired value. The
contribution of the constraint term to the energy is subtracted to
obtain the binding energy of the system at the value of the quadrupole
moment obtained in the calculation. The consistency of the method can be
verified by calculating the ground state energy of a spherical nucleus
as a function of the quadrupole deformation $\beta$. 
For the case of $^{48}$Ca, we find that the ground state energy
reaches its minimum at $\beta=0$, where the nucleus is
spherical, just as we would expect. We also find a small difference of about
0.5 MeV between the calculations for the two values of
$c_{pair}$. This difference is consistent with the small difference in the
pairing energy and the small, but nonzero, average pairing gap found
in the calculations.

\begin{figure}[htbp]
\begin{center}
\epsfig{file=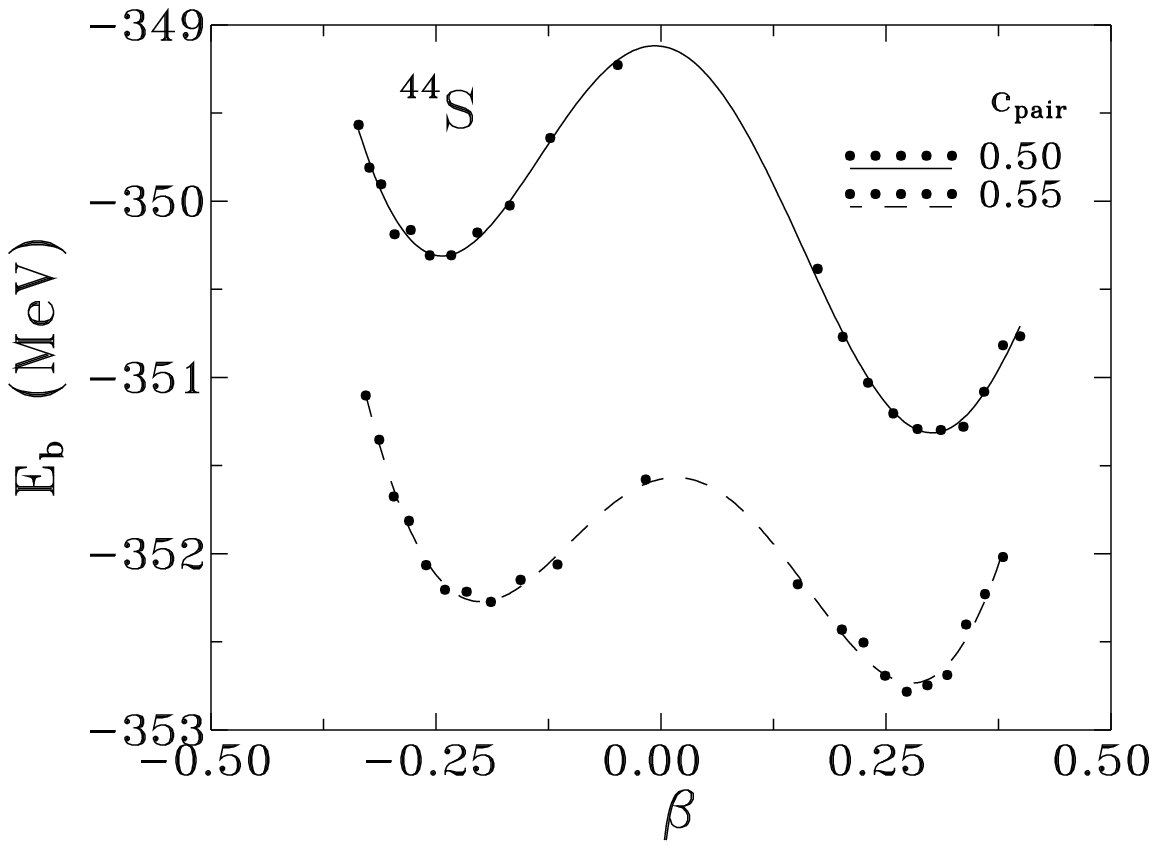,height=7.0cm}
\caption{  Binding energy of $^{44}$S, as a function of the quadrupole
deformation $\beta$, for two values of the parameter $c_{pair}$.
\label{fig14}}
\end{center}
\end{figure}

In Fig.~\ref{fig14}, we present the dependence on the deformation
$\beta$ of the binding energy of the nucleus $^{44}$S, for the two values
of $c_{pair}$. (The energy $E_b$ displayed in the figure actually differs
by a sign from the binding energy,
$E_b(Z,A)=-B(Z,A)$.) Here the variation of $c_{pair}$ from 0.50
to 0.55 makes a difference of almost 3 MeV in the binding energy, in
contrast with the small difference of about 0.5 MeV found for $^{48}$Ca.
The energy curves
for $^{44}$S possess two well defined minima, with the oblate minimum
about 1 MeV above the prolate one in the case of $c_{pair}$=0.50 and
about 0.5 MeV above the prolate minimum when $c_{pair}$=0.55. The
increase in the parameter $c_{pair}$ tends to flatten the energy
curve, thereby diminishing the differences between its peaks and
valleys. The increase in $c_{pair}$ also reduces the value of the
deformation at the minima. The deformation at the oblate
minimum decreases in magnitude by almost 20\% with the increase of
$c_{pair}$ from 0.50 to 0.55 , while the deformation at the prolate
minimum is reduced by about 10\% with the same increase in
$c_{pair}$. 

The experimental value of the quadrupole deformation of $^{44}$S is
$\beta$= 0.258(36)\cite{Gla97,Gla98}, which is in good agreement with
the calculated value, $\beta$=0.28, using $c_{pair}$=0.55.
The  relativistic calculations of Werner et al. and Lalazissis et al. 
furnish a value of the deformation similar to this one.
The RHB calculation of Ref.\onlinecite{Lal98d} yields an oblate minimum
about 200 keV above the prolate
one, which can be compared to our value of about 0.5 MeV between the two minima
and a value of about 0.8 MeV in the RMF+BCS calculation of
Refs. \onlinecite{Wer94} and \onlinecite{Wer96}.
Nonrelativistic HF calculations furnish a value of $\beta$
that is about half of the experimental one\cite{Wer94,Wer96}.

\begin{figure}[htbp]
\begin{center}
\epsfig{file=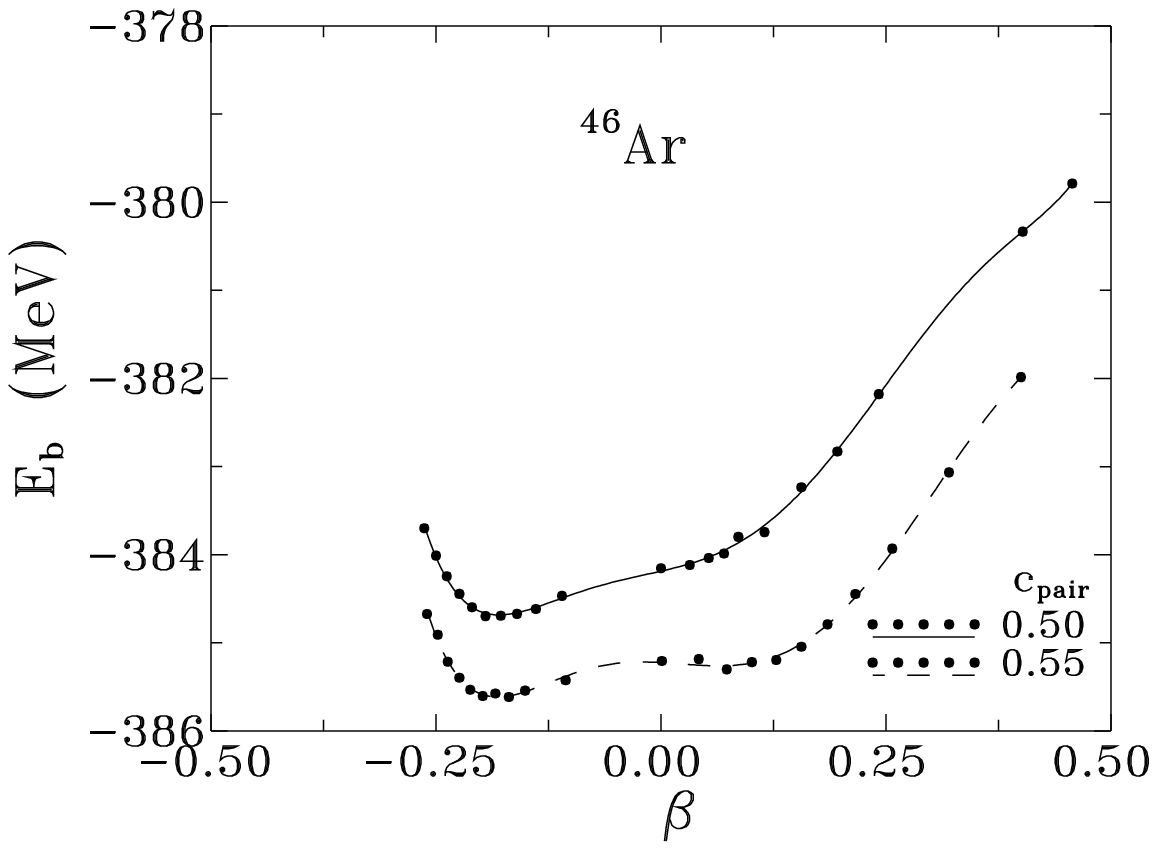,height=7.0cm}
\caption{  Binding energy of $^{46}$Ar, as a function of the quadrupole
deformation $\beta$, for two values of the parameter $c_{pair}$.
\label{fig15}}
\end{center}
\end{figure}

In Fig.~\ref{fig15},  we present the dependence on the deformation
$\beta$ of the binding energy of the nucleus $^{46}$Ar, for two values
of the parameter $c_{pair}$. The difference in the binding energy due
to the variation of $c_{pair}$ is about 1 MeV. The minima of the curves are
much less defined here than they are in the case of $^{44}$S. In
fact, the calculation using $c_{pair}$=0.50 appears to have only an
oblate minimum. The calculation with $c_{pair}$=0.55 possesses two
minima, with the oblate one approximately 200 keV below the prolate one.
Due to the lack of structure in the curves, it is difficult to
determine the reduction of the deformation at the
minima due to the increase in the parameter $c_{pair}$. The
experimental value of the quadrupole deformation of $^{46}$Ar is
$|\beta|$= 0.176(17)\cite{Sch96}, with the sign of the deformation
undetermined experimentally. Assuming the deformation to be oblate, the 
calculated value of $\beta$=-0.18 is in excellent agreement with the data. 
A nonrelativistic HF calculation furnishes results in reasonable agreement
with ours, while its companion RMF+BCS calculation predicts the oblate 
minimum of the $^{46}$Ar energy curve to lie slightly above a spherical
ground state\cite{Wer94,Wer96}.
The RHB calculation using the Gogny interaction yields a flat energy curve
with a barely discernible minimum at $\beta\approx$-0.15\cite{Lal98d}.

We have performed calculations of the binding energy as a function of the 
quadrupole deformation for othet $N=28$ isotones.
The binding energies of $^{40}$Mg and $^{42}$Si are also found to possess both
an oblate and a prolate minimum. Just as for $^{44}$S, we find
that increasing the value of $c_{pair}$ tends to flatten their energy
curves and to reduce the magnitude of the deformation at the minima. The
calculation using $c_{pair}$=0.55 yields, for $^{42}$Si,
a sharp oblate minimum at $\beta$=-0.32, about 1 MeV below a shallow
prolate minimum at $\beta$=0.25. For $^{40}$Mg, the calculation using
$c_{pair}$=0.55 yields a fairly shallow prolate minimum at
$\beta$=0.38, about 0.5 MeV below a shallow oblate minimum at $\beta$=-0.25.
As far as we know, the deformations of these nuclei have not been measured.
As mentioned before, a small neutron pairing gap persists in our calculation
of the spherical nucleus $^{48}$Ca. This gap disappears in the heavier
$N$=28 isotones, $^{50}$Ti, $^{52}$Cr, and $^{54}$Fe, which are also spherical,
but which possess instead an average proton gap.
In Table I, we provide a summary of the ground state energies, deformations,
and average gap parameters of the nuclei in the $N=28$ chain, obtained with
$c_{pair}$=0.55. We note that, except for the lightest nuclei in chain, 
$^{38}$Ne and $^{40}$Mg, the calculations are in good agreement with the
Audi-Wapstra and M\"oller-Nix systematics.

In agreement with the other calculations discussed above, we find the 
even-even nuclei with $A<48$ in the $N=28$ isotonic chain to display the
characteristics of shape coexistence. The nuclei vary between oblate and
prolate ground state deformations and possess both prolate and oblate minima
that are very similar in energy. The values we obtain for the ground state
deformations are very similar to those of the other calculations. The values
obtained for the pairing gaps are different however. The RMF+BCS calculations
of Refs.\onlinecite{Wer94} and \onlinecite{Wer96} fixed the gap at either
75 keV or at 500 keV, values which are relatively small for such light nuclei.
The RHB calculations of Ref.\onlinecite{Lal98d}, which are the most similar
to our calculations in terms of
method, obtained average values of the pairing gap that are about 40\% larger
than ours, but values for the deformations that are almost identical to ours.
If we were to increase the magnitude of the pairing field in our calculations,
the resulting deformations would be smaller than those of the RHB (and other)
calculations. This discrepancy is an indication of the relativistic
effects of the Dirac pairing field that were discussed in the Introduction.
There, we suggested that the Dirac structure of the pairing field should result
in its being more localized on the nuclear surface and, thus, more effective
in limiting the deformation. As the Dirac structure of the
pairing field is the principal difference between our calculations and the RHB
ones, our results lend support to this argument.

\subsection{Deformed nuclei in the $Z\approx 40$ region}

Nuclei in the region of the $Z\approx 40$ subshell closure present
interesting variations in deformation as they deviate from the neutron
magic number $N=50$ on either side of the stability line. Studies of
their level schemes and lifetimes
have shown that the ground-state properties of these nuclei are very
sensitive to small changes in the proton and neutron number\cite{Buc90,Kei95}.

Strontium isotopes present two regions ($N\approx 38$ and $N\approx
60$) of very strong deformations, as large as $\beta=0.4$. This behavior
can be seen in the E2+ measurements as well as in the isotope shifts. The
rapid transition from a spherical shape near the $N$=50 shell closure
to a strongly deformed one can be explained by the
reinforcement of the shell gaps in the single-particle levels for protons
and neutrons \cite{Ham89}.

For krypton isotopes, on the other hand, the expected change in shape
at $N=60$, which occurs for most elements in the $Z=40$ region, does not
appear. Instead, the isotope shifts show a pronounced slope change at $N=50$.
The fact that a strongly deformed ground state does not appear at $N=60$
has been attributed to a reduced proton-neutron interaction at $Z=36$.

Theoretical studies of these nuclei have been carried out using both
relativistic and non-relativistic frameworks. Bonche et al.\cite{Bon85,Bon91}
studied these isotopes extensively including triaxial deformations in the
non-relativistic Hartree-Fock with a Skyrme SIII force, which resulted in a
linear trend for the isotope shifts. They also used the method of generator
coordinates for Sr isotopes and predicted a transition from a spherical to
a deformed shape. However, the isotope shifts could not be reproduced in
any of those calculations.

Many studies using the relativistic mean field approach have been performed in
this region. Early calculations employed the parameter set NL1, which has a
very large asymmetry energy and does not describe the isotope shifts well
\cite{Hir93,She93}. In Ref.~\onlinecite{Lal95}, however, Lalazissis and Sharma
 successfully
described the ground state properties of the Sr and Kr isotopes, by applying
the RMF approach with the NLSH parameter set for the mean field Lagrangian
and the BCS formalism for the pairing correlations. The deviations of the
binding energies from the experimental data were the on order of 0.5\% and
the deformation parameters $\beta$ were in good agreement with the $%
\beta$ values extracted from B(E2) measurements.
It is therefore of interest to perform calculations of the Sr and Kr isotopes
using the Dirac-Hartree-Bogoliubov approach in those regions and compare the
results with the experimental data.

\begin{figure}[htbp]
\begin{center}
\epsfig{file=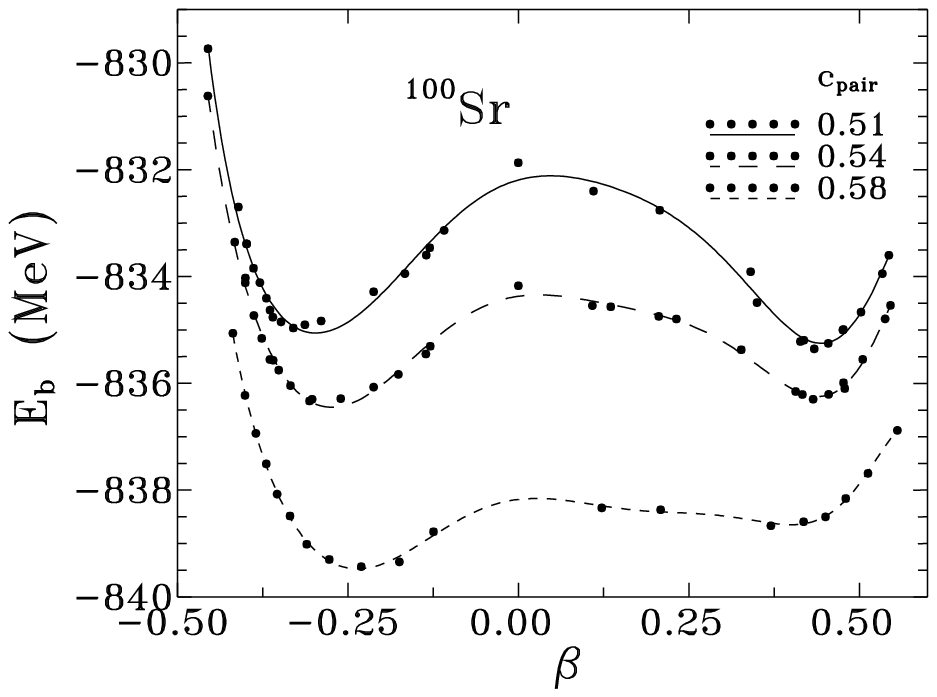,height=7.0cm}
\caption{ Binding energy of $^{100}$Sr, as a function of the quadrupole
deformation $\beta$, for three values of the parameter $c_{pair}$.
\label{fig16}}
\end{center}
\end{figure}

We begin by studying the dependence of the binding energy on the
quadrupole deformation parameter, $\beta$, using the method described
in the previous section. In Fig.~\ref{fig16}, we present the total
binding energy as a function of the deformation $\beta$ of the nucleus
$^{100}$Sr, for three values of the parameter $c_{pair}$. The most singular
difference between the three curves is the change of the absolute minimum,
from a prolate shape in the case of $c_{pair}=0.51$ to an oblate one for
$c_{pair}=0.54$, remaining oblate for $c_{pair}=0.58$. The energy difference
between the two minima is quite small and the oblate minimum becomes deeper
relative to the prolate one as $c_{pair}$ increases. For $c_{pair}=0.51$,
the prolate minimum is 379 keV deeper than the oblate one. For
$c_{pair}=0.54$, the oblate minimum is 17 keV deeper,
while for $c_{pair}=0.58$, it is 823 keV deeper. If we take the oblate
solution as the absolute minimum, the value of the deformation
parameter agrees well with the experimental value of $\beta=0.372\pm
0.008$.

We performed calculations for possible minima of both prolate and oblate
shapes of most of the known even nuclei in the Kr and Sr isotopic chains.
We found many nuclei for which the two minima have very similar binding
energies, an indication of shape coexistence or of a triaxial ground state
deformation.

\begin{figure}[htbp]
\begin{center}
\epsfig{file=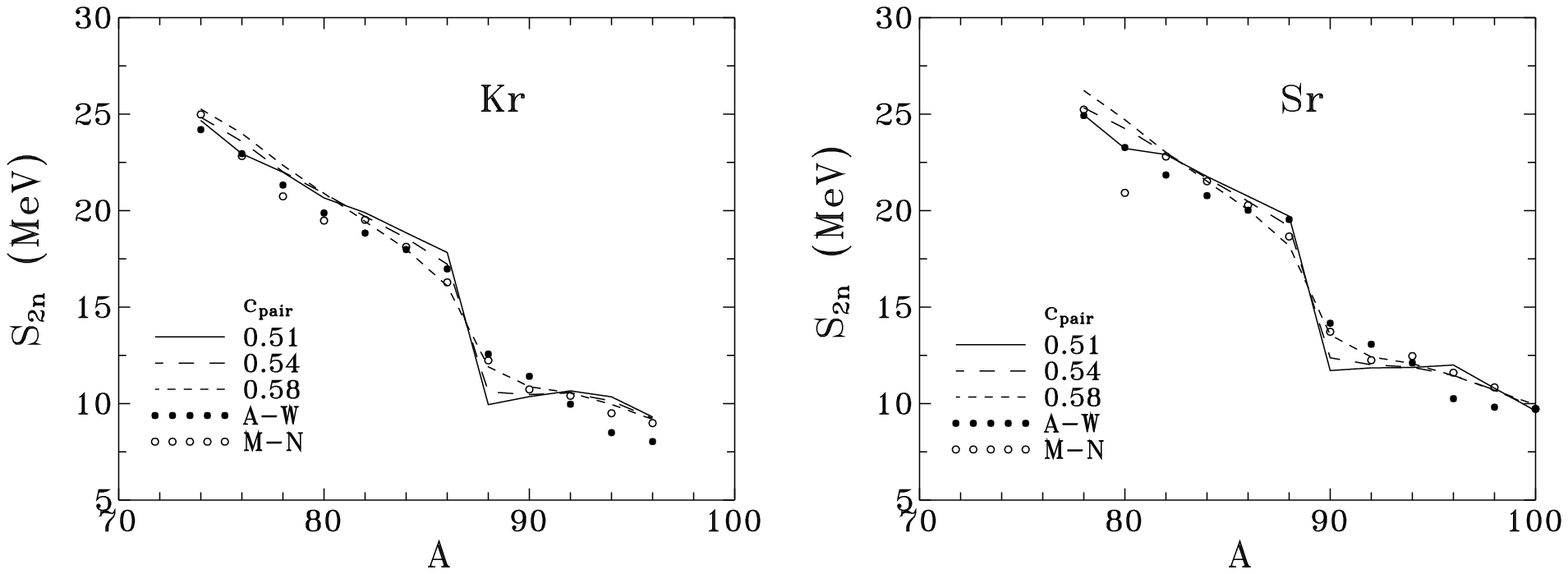,height=6cm}
\caption{ Two neutron separation energy of the even isotopes of Kr and Sr,
as a function of the mass number $A$, for three values of the parameter
$c_{pair}$. The values obtained from the compilation
of experimental masses of Audi and Wapstra\protect\cite{Aud95} (solid circles)
and the M\"oller-Nix systematics\protect\cite{Mol95} (open circles)
are also shown.
\label{fig17}}
\end{center}
\end{figure}

The two-neutron separation energies of the even Kr and Sr isotopes, as
functions of the mass number, are shown in Fig.~\ref{fig17},
for three values of $c_{pair}$.
To prepare this figure, we have selected the solution, among the multiple
minima of each isotope, that has the deepest (absolute) minimum. We compare
our results with the M\"oller-Nix systematics and the Audi-Wapstra
compilation. We find the calculations to agree quite well with both of these.
As in the previous cases, the calculated two-neutron separation energies
are fairly insensitive to the parameter $c_{pair}$.

\begin{figure}[htbp]
\begin{center}
\epsfig{file=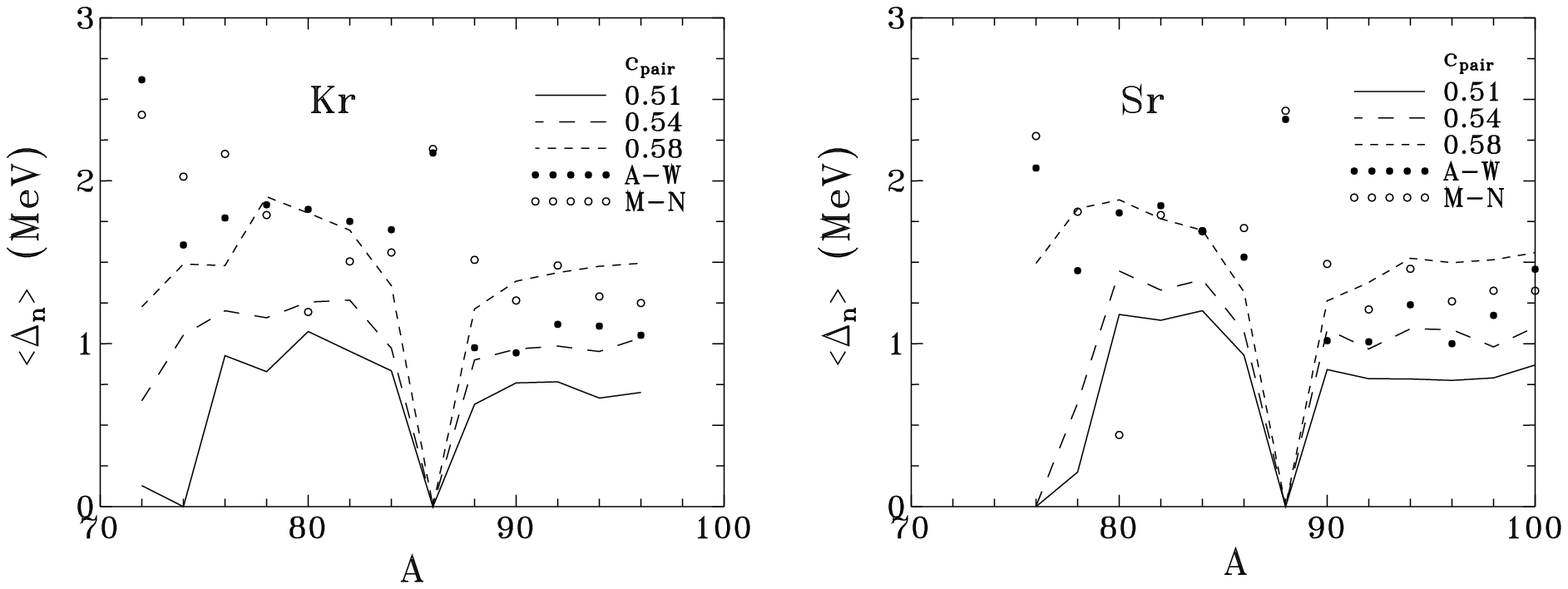,height=6cm}
\caption{ Mean value of the neutron gap parameter of the even
isotopes of Kr and Sr, as a function of the mass number $A$, for three
values of the parameter $c_{pair}$ (lines).  The odd-even mass
differences obtained from the compilation of experimental masses
of Audi and Wapstra\protect\cite{Aud95} (solid circles) and from
the M\"oller-Nix systematics\protect\cite{Mol95} (open circles) are
also plotted.
\label{fig18}}
\end{center}
\end{figure}

In Fig.~\ref{fig18}, we display the mean value of the neutron gap parameter
$<\Delta_{n}>$ of the even mass Kr and Sr isotopes, as a function of the mass
number, for the same three values of $c_{pair}$. We find the best agreement
with the M\"oller-Nix and Audi-Wapstra values to be given by the calculations
using $c_{pair}=0.58$. The shell closure at $N=50$ is clearly visible as a
zero in the calculated pairing gap of both isotopoic chains.

In Table II, we present the results for the total binding energy and the
deformation parameter $\beta$ obtained in the present calculations,
for $c_{pair}$=0.58, together with the experimental data. The value
$c_{pair}$=0.58 furnishes the best description of the total
binding energies and of the mean value of the neutron pairing gap. 
The cases in which the difference between two minima is less than 200
keV are marked with a star and the results for the two deformations are
presented. In general, the calculated binding energies agree quite well
with the experimental values, although the proton-rich isotopes are
slightly underbound in the calculations.

Our calculations, with one exception,
predict that the non-spherical even nuclei in the two isotopic chains will
have an oblate ground state deformation. This contrasts sharply with the 
M\"oller-Nix systematics which, with two exceptions, predicts prolate
deformations for the ground states of the non-spherical nuclei of the
two chains. Such disagreement is not too surprising, given the sensitivity
to the interaction parameters of the binding energy versus deformation curve
of many of these nuclei. An example of this sensitivity was displayed in
Fig.~\ref{fig16}, in which the ground state of $^{100}$Sr was seen to be
prolate for $c_{pair}=0.51$ but oblate for $c_{pair}=0.54$ and 0.58. We note
that, of the the five nuclei for which we found a second minimum within 200 keV
of the ground state one, the deformation of the second minimum of four of
these nuclei is in agreement with the M\"oller-Nix prediction. Experimentally,
large ground state deformations of $\beta\approx0.44$ have been found in
$^{76}$Sr, $^{74}$Kr, and $^{76}$Kr\cite{Ham89}. These values are in agreement
with the M\"oller-Nix ones and with our result for $^{76}$Sr. They are not
consistent with our results for the Kr isotopes.

\begin{figure}[htbp]
\begin{center}
\epsfig{file=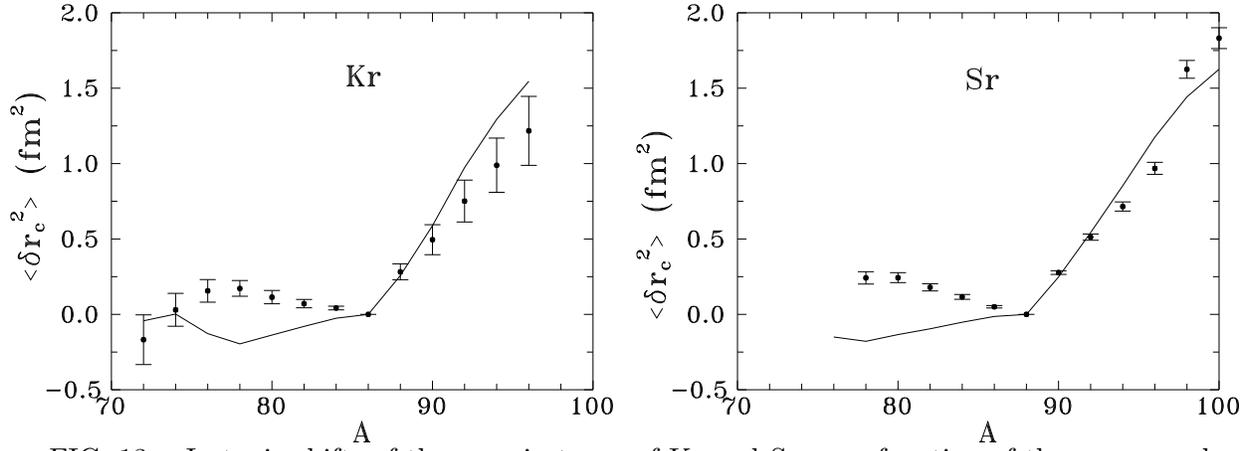,height=5.9cm}
\caption{   Isotopic shifts of the even isotopes of
Kr and Sr, as a function of the mass number $A$, for $c_{pair}$=0.58 (lines). 
The experimental values are displayed as points with error bars. 
\label{fig20}}
\end{center}
\end{figure}

The calculated isotope shifts of the even-even Kr and Sr isotopes,
with respect to the reference nuclei $^{86}$Kr and $^{88}$Sr,
respectively, are compared to the experimental data in Fig.~\ref{fig20}.
We display only the results for the deepest energy minimum of each
isotope in the figure. In the cases in which a second minimum is close in
energy to the deepest one, their isotopic shifts are also very similar.
We find the agreement with the data for the neutron rich side to be
very good. In addition, the positions of the kink due to the shell
effect for the Kr and Sr isotopes are also well reproduced. However, the
calculations underestimate the isotope shift on the proton rich side of 
the stability line.

\section{Summary}

We have used an extension of the Gorkov formalism for the description
of pairing to develop a 
Dirac-Hartree-Fock-Bogoliubov approximation to the ground state wave function
and energy of finite nuclei. We have applied it to spin-zero proton-proton and
neutron-neutron pairing within the Dirac-Hartree-Bogoliubov approximation
(we neglected the Fock term). We have retained the full Dirac structure of the
pairing field that is permitted by the symmetries of the problem. We find
the Dirac pairing structure to be dominated by a scalar term and the
zero component of a vector term, as has also been found to be the case for
$^1$S$_0$ pairing in symmetric nuclear matter. 

In our calculations, we use a zero-range approximation to the pairing
interaction, which results in a local pairing field.
We justified the zero-range approximation by arguing that the effective
length for spatial variations of the wavefunctions, in the calculations
performed at present, is larger than the range of the nonlocality of the 
pairing interaction, rendering the effects of the nonlocality close to
negligible.

We studied the effects of the Dirac pairing field on the properties
of the even-even nuclei of the isotopic chains of Ca, Ni and Sn (spherical)
and Kr and Sr
(deformed), as well as the $N$=28 isotonic chain. We first
studied the isotopic chains of the spherical nuclei in order to determine the
sensitivity of various nuclear observables to the interaction in the pairing
channel. We found the two-neutron separation energies,
root-mean-square radii and isotopic shifts of the spherical nuclei
to be fairly independent of the strength of the pairing interaction.
The binding energy, on the contrary, is quite sensitive to the
pairing interaction strength (at least when the pairing is not
identically zero due to a shell closure). Quite obviously, the
average pairing gap also displays a strong dependence on
the strength of the pairing interaction. We observed that an adequate
choice of this strength simultaneously
yields reasonable values for the binding energy, near the
stability line, and rough agreement between the average pairing gap
and the experimental gap, defined in terms of even-odd mass differences.
Far from the stability line, our calculated binding energies showed
increasing discrepancies with the experimental data, suggesting that
the NL3 parameter set might require further adustment, at least when used
in a DHB formalism. All in all, however, our results show good agreement 
with the data and with previous calculations.

To study the deformed Kr and Sr nuclei and the deformed nuclei of the
$N$=28 isotonic chain, we performed a sequence of constrained
Dirac-Hartree-Bogoliubov
calculations, as a function of the quadrupole deformation $\beta$,
for each of the nuclei in question. Each sequence of calculations
provided us with a curve of binding energy versus deformation that
permitted the localization of the equilibrium configurations of the
nucleus. Most of the nuclei in these three chains present both a
prolate and an oblate minimum (and, at times, a slight spherical
minimum as well). In agreement with other calculations, we found
nuclei in each of the chains for which the energies of the oblate and
prolate minima were very similar, raising the possibility of shape
coexistence or triaxial ground state deformations. For the most part, our
results present reasonable agreement with the data and with previous
calculations.  We did, however, observe a discrepancy between our average
pairing gaps and those of the RHB calculations of Ref.\cite{Lal98d}, which
we took as an indication of the relativistic suppression of
the pairing field due to its Dirac structure. Our DHB calculations furnished
quadrupole deformations similar to the RHB ones for average values of
the pairing gap that are about 30\% smaller than theirs. 

Based on the nuclei we have studied here, we conclude that the
DHB approximation can provide a description of the binding energies, rms
radii and ground state deformations of finite nuclei that is at
least as good as that provided by the nonrelativistic HFB approximation
or by the RHB approximation. Our claim that the DHB approximation can
provide a more reliable description than the others, however, has not been
demonstrated here. This will require further study of the parameters that
enter the pairing interaction, in order to better fix their values.

In the future, we plan to extend our calculations to a larger set of
isotopic chains and to analyze in detail their proton- and neutron-rich
tails, in order to obtain a set of parameters that describes these better. 
We plan to extend our calculations to odd nuclei by
including a blocking term. As the blocking term varies,
along with the other states, during the search for the Dirac-Hartree-Bogoliubov
stationary point, this does not seem to be a trivial task. In order to
facilitate it, we plan to first develop an extended BCS approximation, 
based on the self-consistent Hartree eigenstates of the mean field
in conjunction with the relativistic pairing field. Such an approximation
could also serve as a faster first stage to the Dirac-Hartree-Bogoliubov
calculations, which are quite time-consuming at the present.
Eventually, we also intend to include the nonlocality of the pairing
interaction, in order to use more realistic interactions containing the
relatively long-range effects of pion exchange.

\section*{Acknowledgments} 
The authors would like to thank P. Ring for providing them
with an early version of the Munich deformed relativistic Hartree code\cite
{Pos97b}, which served as the starting point for the code used for the 
calculations described
here. The authors acknowledges partial support from FAPESP
(Funda\c{c}\~{a}o de Amparo \`{a} Pesquisa do Estado de S\~{a}o Paulo).
B.V.C. acknowledges partial support from the CNPq (Conselho
Nacional de Pesquisa e Desenvolvimento).

\appendix

\section{Time-reversed pairs of states}

We begin by writing the complex conjugate of Eq.~(\ref{autoeq1}) as

{\footnotesize
\begin{eqnarray}
\int d^{3}y\left(
\begin{array}{cc}
\gamma _{0}((\varepsilon +\mu )\delta (\vec{x}-\vec{y})-Bh^{*}(\vec{x},\vec{y%
})B^{\dagger }) & B\Delta ^{*}(\vec{x},\vec{y})B^{\dagger } \\
B\bar{\Delta}^{*}(\vec{x},\vec{y})B^{\dagger } & ((\varepsilon -\mu ) \delta
(\vec{x}-\vec{y})+Bh_{T}^{*}(\vec{x},\vec{y})B^{\dagger })\gamma _{0}
\end{array}
\right) \left(
\begin{array}{c}
BU^{*}(\vec{y}) \\
BV^{*}(\vec{y})
\end{array}
\right) = 0\,,
\end{eqnarray}
} where $B=\gamma _{5}C$ is the Dirac matrix part of the time-reversal
operator. Analyzing the Hamiltonian term, we see that
\begin{eqnarray}
B\,h^{*}(\vec{x},\vec{y})B^{\dagger } &=&B\,(i\vec{\alpha}^{*}\cdot \vec{%
\nabla}+\beta ^{*}M)B^{\dagger }\delta (\vec{x}-\vec{y})+B\,\beta ^{*}\Sigma
^{*}(\vec{x},\vec{y})B^{\dagger } \\
&=&(-i\vec{\alpha}\cdot \vec{\nabla}+\beta M)\delta (\vec{x}-\vec{y}%
)+B\,\beta ^{*}\Sigma ^{*}(\vec{x},\vec{y})B^{\dagger }.  \nonumber
\end{eqnarray}
The self-energy term, in turn, can be put into the form {\footnotesize
\begin{eqnarray}
B\,\beta ^{*}\Sigma (\vec{x},\vec{y})B^{\dagger } &=&\delta (\vec{x}-\vec{y}%
)\sum_{j}B\,\gamma _{0}^{*}\Gamma _{j\alpha }^{*}(\vec{x})B^{\dagger }\int
d^{3}z\,D_{j}^{\alpha \beta }(\vec{x}-\vec{z})\sum_{\varepsilon _{\gamma
}<0}U_{\gamma }^{T}(\vec{z})B^{\dagger }B\gamma _{0}^{*}\Gamma _{j\beta
}^{*}(\vec{z})B^{\dagger }BU_{\gamma }^{*}(\vec{z})B^{\dagger }  \nonumber \\
&&\qquad \qquad -\sum_{j}B\gamma _{0}^{*}\Gamma _{j\alpha }^{*}(\vec{x}%
)B^{\dagger }D^{\alpha \beta }(\vec{x}-\vec{y})\sum_{\varepsilon _{\gamma
}<0}BU_{\gamma }^{*}(\vec{x})U_{\gamma }^{T}(\vec{y})B^{\dagger }B\,\gamma
_{0}^{*}\Gamma _{j\beta }^{*}(\vec{y})B^{\dagger }\,.
\end{eqnarray}
}

If we now look closely at the vertex functions we are considering here, we
find

for the $\sigma $ meson: $B\gamma _{0}^{*}1B^{\dagger }\ldots $ $B\gamma
_{0}^{*}1B^{\dagger }=\gamma _{0}1\ldots \gamma _{0}1,$

for the $\omega $ meson: $B\gamma _{0}^{*}\gamma _{\mu }^{*}B^{\dagger
}g^{\mu \nu }\ldots $ $B\gamma _{0}^{*}\gamma _{\nu }^{*}B^{\dagger }=\gamma
_{0}\gamma _{\mu }\,g^{\mu \nu }\ldots \gamma _{0}\gamma _{\nu },$

for the $\rho $ meson: $B\gamma _{0}^{*}\gamma _{\mu }^{*}\tau
_{i}^{*}B^{\dagger }g^{\mu \nu }\delta ^{ij}\ldots $ $B\gamma _{0}^{*}\gamma
_{\nu }^{*}\tau _{j}^{*}B^{\dagger }=\gamma _{0}\gamma _{\mu }\tau
_{i}\,g^{\mu \nu }\delta ^{ij}\ldots \gamma _{0}\gamma _{\nu }\tau _{j},$

for the photon: $B\frac{\left( 1+\tau _{3}^{*}\right) }{2}\gamma
_{0}^{*}\gamma _{\mu }^{*}B^{\dagger }g^{\mu \nu }\ldots $ $B\frac{\left(
1+\tau _{3}^{*}\right) }{2}\gamma _{0}^{*}\gamma _{\nu }^{*}B^{\dagger }=%
\frac{\left( 1+\tau _{3}\right) }{2}\gamma _{0}\gamma _{\mu }\,g^{\mu \nu
}\ldots \frac{\left( 1+\tau _{3}\right) }{2}\gamma _{0}\gamma _{\nu }.$ In
short, the products of the pairs of vertices remain unchanged by the
transformation. We thus have for the transformed self energy, {\footnotesize
\begin{eqnarray}
B\beta ^{*}\,\Sigma ^{*}(\vec{x},\vec{y})B^{\dagger } &=&\delta (\vec{x}-%
\vec{y})\sum_{j}\gamma _{0}\Gamma _{j\alpha }(\vec{x})\int
d^{3}z\,D_{j}^{\alpha \beta }(\vec{x}-\vec{z})\left( \sum_{\varepsilon
_{\gamma }<0}U_{\gamma }^{T}(\vec{z})B^{\dagger }\gamma _{0}\Gamma _{j\beta
}(\vec{z})BU_{\gamma }^{*}(\vec{z})\right) \\
&&\qquad \qquad -\sum_{j}\gamma _{0}\Gamma _{j\alpha }(\vec{x})D^{\alpha
\beta }(\vec{x}-\vec{y})\left( \sum_{\varepsilon _{\gamma }<0}BU_{\gamma
}^{*}(\vec{x})U_{\gamma }^{T}(\vec{y})B^{\dagger }\gamma _{0}\right) \Gamma
_{j\beta }(\vec{y})\,,  \nonumber
\end{eqnarray}} 
\newline 
in which the only differences from the original expression for the self
energy are the propagator contributions, written in terms of the solutions
of the Dirac-Gorkov equation, that enter the expression.

We can perform the same analysis on the pairing term, for which we find a
similar result,

\begin{equation}
B\,\Delta ^{*}(\vec{x},\vec{y})B^{\dagger }\;=-\sum_{j}\Gamma _{j\alpha }(%
\vec{x})D_{j}^{\alpha \beta }(\vec{x}-\vec{y})\left( \sum_{\varepsilon
_{\gamma }<0}BU_{\gamma }^{*}(\vec{x})V_{\gamma }^{T}(\vec{y})B^{\dagger
}\gamma _{0}\right) A\Gamma _{j\beta }^{T}(\vec{y})A^{\dagger }\;,
\end{equation}
in which the only difference between the original expression and the
transformed one is again the propagator's contribution to the expression.
Analyses of the other terms, $B\,h_{T}^{*}(\vec{x},\vec{y})B^{\dagger }$ and
$B\,\bar{\Delta}^{*}(\vec{x},\vec{y})B^{\dagger }$, yield similar results.

We can thus show that the transformed wave vector {\footnotesize $\left(
\begin{array}{l}
BU^{*}(\vec{x}) \\
BV^{*}(\vec{x})
\end{array}
\right) $} satisfies a similar equation

{\footnotesize
\begin{eqnarray}
\int d^{3}y\left(
\begin{array}{cc}
\gamma _{0}((\varepsilon +\mu )\delta (\vec{x}-\vec{y})-h(\vec{x},\vec{y}))
& \Delta (\vec{x},\vec{y}) \\
\bar{\Delta}(\vec{x},\vec{y}) & ((\varepsilon -\mu )\delta (\vec{x}-\vec{y}%
)+h_{T}(\vec{x},\vec{y}))\gamma _{0}
\end{array}
\right) \left(
\begin{array}{c}
BU^{*}(\vec{y}) \\
BV^{*}(\vec{y})
\end{array}
\right) = 0\,,
\end{eqnarray}
} to that satisfied by the wavevector {\footnotesize $\left(
\begin{array}{l}
U(\vec{x}) \\
V(\vec{x})
\end{array}
\right) $} . Each is a solution of an equation with the same energy
eigenvalue but with orbital quantum numbers and mean fields that have the
time-reversed Dirac structure of the other. If each member of these pairs of
states possessing time-reversed Dirac structure are equally occupied,
the sums over states that enter into the definition of the self-energy and
pairing fields will be invariant under the transformation that
temporally inverts the Dirac structure. That is, we then have that

\begin{equation}
\sum_{\varepsilon _{\gamma }<0}U_{\gamma }(\vec{x})U_{\gamma }^{\dagger }(%
\vec{y})\ =\sum_{\varepsilon _{\gamma }<0}BU_{\gamma }^{*}(\vec{x})U_{\gamma
}^{T}(\vec{y})B^{\dagger }
\end{equation}
and
\begin{equation}
\sum_{\varepsilon _{\gamma }<0}U_{\gamma }(\vec{x})V_{\gamma }^{\dagger }(%
\vec{y})\ =\sum_{\varepsilon _{\gamma }<0}BU_{\gamma }^{*}(\vec{x})V_{\gamma
}^{T}(\vec{y})B^{\dagger }.
\end{equation}
In this case, both of the wavevectors {\footnotesize $\left(
\begin{array}{l}
U(\vec{x}) \\
V(\vec{x})
\end{array}
\right) $ }and {\footnotesize $\left(
\begin{array}{l}
BU^{*}(\vec{x}) \\
BV^{*}(\vec{x})
\end{array}
\right) $ }satisfy the same Dirac-Gorkov equation with the same energy
eigenvalue. Then either both or neither of them will enter the propagator's
contributions to the self-consistency equations and these contributions will
indeed be invariant under time-inversion of their Dirac structure. We
assume this to be the case.

\newpage

\begin{center}
{\bf Table 1}
\end{center}

\[
\begin{tabular}{|c|ccc|ccc|c|c|}
\hline
  &  & B.E. &  & &$\beta$ &  & $<\Delta_n>$ & $<\Delta_p>$ \\
  $\qquad\qquad$ &$\quad$DHB$\quad$&$\quad$A-W$\quad$&$\quad$M-N$\quad$&
$\quad$M-N & & DHB$\quad$&$\quad$DHB$\quad$&$\quad$DHB$\quad$\\
\hline
$^{38}$Ne & -225.19 &         & -214.03 & -0.29 &  & 0.28 & 2.21 & 1.19 \\
$^{40}$Mg & -272.48 &         & -268.04 & -0.29 &  & 0.39 & 1.86 & 0.00 \\
$^{42}$Si & -315.53 & -313.04 & -315.16 & -0.32 & & -0.32 & 1.24 & 0.00 \\
$^{44}$S  & -352.78 & -353.49 & -351.99  & 0.00 &  & 0.28 & 1.42 & 0.00 \\
$^{46}$Ar & -385.58 & -386.91 & -386.18  & 0.00 & & -0.18 & 1.19 & 0.00 \\
$^{48}$Ca & -415.05 & -415.98 & -415.56  & 0.00 &  & 0.00 & 0.84 & 0.00 \\
$^{50}$Ti & -436.99 & -437.77 & -438.65  & 0.00 &  & 0.00 & 0.00 & 1.55 \\
$^{52}$Cr & -455.58 & -456.34 & -457.27  & 0.00 &  & 0.00 & 0.00 & 1.52 \\
$^{54}$Fe & -470.69 & -471.75 & -472.58  & 0.00 &  & 0.00 & 0.00 & 1.06 \\
$^{56}$Ni & -482.67 & -483.98 & -484.48  & 0.00 &  & 0.00 & 0.00 & 0.00 \\
 \hline
\end{tabular}
\]

\newpage

\begin{center}
{\bf Table 2}
\end{center}

{\small \[
\begin{tabular}{|ccccc|ccccc|}
\hline
   &                  &Kr      &            &     &    &                  &
Sr   &             & \\  \hline
   & $\qquad\qquad$  $B.E.$         &        &$\qquad\qquad$$\beta$&
&    &
   $\qquad\qquad$$B.E.$           &         &$\qquad\qquad$$\beta$ & \\
$A$& $DHB$            &  $Exp.$& $DHB$      &$M-N$&$A$ & $DHB$            &
$Exp.$  & $DHB$       & $M-N$\\ \hline
72 &-604.25           &-607.11 &-0.32       &-0.35& 76 &-635.48           &
-638.08 & 0.49        &0.42\\
74 &-629.51           &-631.28 &-0.30       & 0.40& 78 &-661.70$^{*}$     &
-663.01 & -0.17 (0.47)&0.42 \\
76 &-653.51$^{*}$     &-654.23 &-0.20 (0.00)& 0.40& 80 &-686.41           &
-686.28 & 0.00        &0.05 \\
78 &-675.86           &-675.55 & 0.00       &-0.23& 82 &-709.43           &
-708.13 & 0.00        &0.05 \\
80 &-696.75           &-695.43 & 0.00       & 0.06& 84 &-730.94           &
-728.90 & 0.00        &0.05 \\
82 &-716.19           &-714.27 & 0.00       & 0.07& 86 &-750.99           &
-748.92 & 0.00        &0.05 \\
84 &-734.19           &-732.26 & 0.00       & 0.06& 88 &-769.17           &
-768.46 & 0.00        &0.05\\
86 &-750.29           &-749.23 & 0.00       & 0.05& 90 &-782.74           &
-782.63 & 0.00        &0.05\\
88 &-762.20           &-761.80 & 0.00       & 0.06& 92 &-795.15$^{*}$     &
-795.75 &-0.12 (0.10) &0.08\\
90 &-773.08$^{*}$     &-773.21 &-0.14 (0.14)& 0.16& 94 &-807.22           &
-807.81 &-0.18        &0.26 \\
92 &-783.63$^{*}$     &-783.22 &-0.27 (0.20)& 0.23& 96 &-818.67           &
-818.10 &-0.23        &0.34\\
94 &-793.59           &-791.76 &-0.29       & 0.31& 98 &-829.36           &
-827.87 &-0.26        &0.36\\
96 &-802.77           &-799.95 &-0.16       & 0.34&100 &-839.29           &
-837.62 &-0.25        &0.37\\ \hline
\end{tabular}
\]}

\end{document}